\documentclass[12pt,a4paper]{article} 
\pdfoutput=1
\usepackage{jheppub}
\usepackage{graphicx, amsmath, xcolor, adjustbox}
\usepackage[numbers, elide]{natbib}

\def\equationautorefname~#1\null{eq.\,(#1)\null}

\makeatletter\g@addto@macro\bfseries{\boldmath}\makeatother

\DeclareRobustCommand\hbar{{\mathchar'26\mkern-9muh}}

\usepackage{ifpdf,feynmp,graphicx}
\ifpdf\fi
\AtBeginDocument{\begin{fmffile}{fgraph}
\fmfcmd{%
arrow_ang := 11;
arrow_len := 3.5thick;
curly_len := 2.5thick;
style_def top expr p =
  save oldpen;
  pen oldpen;
  oldpen := currentpen;
  pickup oldpen scaled 1.25;
  ccutdraw p;
  pickup oldpen scaled .25;
  cullit; undraw p; cullit;
  cfill (arrow p);
  pickup oldpen scaled 1;
enddef;}
}

\newcommand{\bea}{\begin{eqnarray}}
\newcommand{\eea}{\end{eqnarray}}
\newcommand{\beq}{\begin{equation}}
\newcommand{\eeq}{\end{equation}}

\newcommand{\rat}{r^8_6}%

\begin{document}

\title{Charting the Higgs self-coupling boundaries}

\author[a]{Gauthier Durieux,}
\author[a]{Matthew McCullough}
\affiliation[a]{Theoretical Physics Department, CERN,\\
Esplanade des Particules 1, Geneva CH-1211, Switzerland
}
\author[b]{and Ennio Salvioni}
\affiliation[b]{Dipartimento di Fisica e Astronomia, Università di Padova and INFN, Sezione di Padova,\\
Via Marzolo 8, Padua I-35131, Italy}

\abstract{%
Could new physics first manifest itself in Higgs self-coupling measurements?
In other words, how large could deviations in the Higgs self-coupling be, if other Higgs and electroweak measurements are compatible with Standard Model predictions?
Using theoretical arguments supported by concrete models, we derive a bound on the ratio of self-coupling to single-Higgs coupling deviations in ultraviolet completions of the Standard Model where parameters are not fine-tuned.
Broadly speaking, a one-loop hierarchy is allowed.
We thus stress that self-coupling measurements at the LHC and future colliders probe uncharted parameter space, presenting discovery potential even in the absence of emerging hints in single-Higgs coupling measurements.
For instance, if other observables show less than two-sigma deviations by the end of the LHC programme, the Higgs self-coupling deviations could still exceed $200\%$ in the models discussed, without introducing fine-tuning of ultraviolet parameters.
}

\preprint{CERN-TH-2022-140}

\maketitle

\section{Introduction}
\label{sec:introo}
Defined as the coefficient of the on-shell three-Higgs amplitude (in complex kinematics), the Higgs self-coupling is firmly predicted in the Standard Model (SM), being fixed at tree level by the Higgs mass and Fermi constant.
As with the other Higgs interactions, we still wish to test the SM prediction by measuring the self-coupling at colliders.
This is a fiendishly challenging task, however, making the Higgs self-coupling somewhat of the white whale of Higgs physics.
The HL-LHC will only probe order-one departures from the SM prediction and pushing the energy frontier further will be required to gain an order of magnitude in precision.
In contrast, the Higgs coupling to vector bosons will be measured to the few-percent level by the HL-LHC, and to the per-mille level at proposed next-generation facilities.

Defining the fractional deviations relative to the SM predictions,
\beq
\delta_{h^3}	\equiv \frac{C_{h^3}-C^{\text{SM}}_{h^3}}{C^{\text{SM}}_{h^3}}
~~, \qquad
\delta_{VV}	\equiv \frac{C_{hVV}-C^{\text{SM}}_{hVV}}{C^{\text{SM}}_{hVV}}
~~,
\label{eq:delta-def}
\eeq
a model builder may argue that, in many well-motivated theories beyond the SM (BSM), one finds that $\delta_{VV}$ and $\delta_{h^3}$ are of similar magnitudes.
Hence, a constraint on $\delta_{VV}$ may, in practice, limit the magnitude of $\delta_{h^3}$ one should expect.
On the other hand, a more phenomenologically minded physicist may advance that, in full generality, we have very little information on $\delta_{h^3}$ beyond established direct experimental constraints.
Here, we aim to bridge this potential gap in perspective by studying, with both general theoretical arguments (\autoref{sec:intro}) and explicit models (\hyperref[sec:tree]{sections \ref*{sec:tree}} and \ref{sec:pNGB}), the range of values allowed for the ratio
\beq
\left| \frac{\delta_{h^3}}{\delta_{VV}}  \right| ~~,
\label{eq:ratio}
\eeq
in generic ultraviolet (UV) completions.
In this paper, we call `generic' any UV completion in which no parameters allowed by symmetries have been arbitrarily fine-tuned to achieve a specific value of the Higgs self-coupling.  
We stress that the Higgs naturalness problem is a different, though related, question.
In all models considered here, the Higgs self-coupling is calculable in terms of the fundamental parameters of the theory.
However, this is not always the case for the Higgs mass and vacuum expectation value.
In models where these are not calculable, we ignore the question of Higgs naturalness.

Assuming that new physics is sufficiently heavy to be described by an effective field theory (EFT) at collider energies, we argue in \autoref{sec:intro} that there is indeed an effective upper limit on the $| {\delta_{h^3}}/{\delta_{VV}}|$ ratio in generic UV completions.
Thus one does have some notion for the expected maximum magnitude of self-coupling corrections given an experimental constraint on $\delta_{VV}$.
However, in practice, this limit is relatively weak as it is essentially given by a loop factor $(4 \pi)^2\approx 160$.
An example generic UV completion~\cite{Logan:2015xpa,Chala:2018ari} illustrating that this limit can be saturated is presented in \autoref{sec:tree}.
The existence and weakness of this bound has important consequences for future collider discussions.
By the end of HL-LHC running, if single-Higgs couplings are still measured to be SM-like within experimental errors, then a significant Higgs self-coupling deviation, and hence new physics discovery, is still possible in generic UV theories.

We emphasise that we do not mean `generic' to imply that the UV completion is a commonly studied theory, such as the classic supersymmetric or composite Higgs scenarios.
Therefore, the relatively weak upper bound we find on $|\delta_{h^3}/\delta_{VV}|$ reveals an important limitation to charting the physics landscape of future facilities using solely the canon of BSM models that have been put forward to address outstanding questions in the SM, since such models may not map the full extent of what is theoretically possible for the Higgs self-coupling.

In \autoref{sec:pNGB}, we study this question further in the context of a recently proposed class of pseudo-Nambu-Goldstone boson (pNGB) Higgs models, which can realise natural electroweak symmetry breaking by means of non-minimal symmetry breaking parameters~\cite{Durieux:2021riy,Durieux:2022sgm}.
Here again, working within an EFT description, we find that the ratio in \autoref{eq:ratio} can be significantly enhanced without resorting to fine-tuning.
Furthermore, in this setting, we find that a dimension-6 truncation of the EFT description can fail to capture the physics of the Higgs self-coupling.
This signals a need for caution: in some scenarios, as concerns the Higgs self-coupling, operators of even higher dimension ($>6$) may have significantly enhanced Wilson coefficients, making them equally or more important than the dimension-6 contributions.
We also study the question of perturbative unitarity within this class of models, finding moderate constraints.

This paper is far from the first study of the theoretical landscape of the Higgs boson self-coupling, see for instance \cite{Hollik:2001px,Kanemura:2002vm,Barger:2003rs,Grojean:2004xa,Giudice:2007fh,Gupta:2013zza,Efrati:2014uta,deBlas:2014mba,Azatov:2015oxa,Goertz:2015dba,Dawson:2015oha,Buttazzo:2015bka,Ginzburg:2015yva,Liu:2016idz,DiLuzio:2016sur,Baglio:2016bop,DiVita:2017eyz,DiLuzio:2017tfn,Carvalho:2017vnu,Chang:2019vez,Falkowski:2019tft,Agrawal:2019bpm,Abu-Ajamieh:2020yqi}.
Thus in \autoref{sec:others}, we compare and contrast our approach and results with some recent literature.
Conclusions are presented in \autoref{sec:conclusions}.

\section{General perspective on \texorpdfstring{$\delta_{h^3} / \delta_{VV}$}{delta h3 / delta VV}}
\label{sec:intro}
Consider a UV completion of the SM featuring new states at a mass scale $M$ above the direct reach of current measurements.
At energies below $M$, this theory will imprint itself on higher-dimension SM operators and leave traces in Higgs couplings.
We wish to understand how large of a hierarchy there can be between the Higgs self-coupling modification $\delta_{h^3}$ and other signals of the new physics, such as corrections to the Higgs couplings to weak vector bosons $\delta_{VV}$ and the precision electroweak $\widehat{T}$ parameter.
Given an arbitrarily fine-tuned UV completion, these hierarchies could in principle be arbitrarily large.
However, here we are interested in UV completions which are not fine-tuned for this purpose.

From an EFT perspective, this would at first glance correspond to asking whether a generic UV completion can generate the dimension-$6$ operator
\beq
\mathcal{O}_6 = -\, \frac{1}{M^2} |H|^6 ~~,
\eeq
and also operators such as
\beq
\mathcal{O}_H = \frac{1}{M^2} ( \partial_\mu |H|^2 )^2  ~~,~~ \mathcal{O}_R = \frac{1}{M^2} |H|^2 |D_\mu H|^2 ~~,~~ \mathcal{O}_T = \frac{1}{M^2} |H^\dagger D_\mu H|^2 ~~,
\label{eq:operators}
\eeq
but with coefficients $|c_6| \gg |c_{H,R,T}|$.
The $|H|^4D^2$ operators in~\autoref{eq:operators} affect Higgs couplings to weak gauge bosons and fermions, as well as the $\widehat{T}$ parameter.

\subsection{\texorpdfstring{$\hbar$}{hbar} as a guide}
\label{sec:hbar-counting}
Let us first investigate possibilities based on $\hbar$ counting, which can equivalently be phrased in terms of coupling dimensions (see e.g.\ \cite{Espinosa:2016ovf, Giudice:2016yja}), or na\"ive dimensional analysis weight \cite{Manohar:1983md}.
The Wilson coefficient $c_6$ has four powers of coupling dimension.
Crucially, this is unique within the SMEFT at dimension-$6$.
Motivated by this observation, consider a UV completion with typical mass scale $M$ and a coupling parameter $\kappa$ in the infrared (IR), which carries four powers of coupling dimension and dominates the interactions between the Higgs and new physics.
Importantly we assume $\sqrt[4]{\kappa}v/M<1$ and all dimensionless coefficients to be of order one,\footnote{%
Note that, due to coefficients that are not of order one, operators of dimension higher than six dominate self-coupling corrections in the scenario of \autoref{sec:pNGB}.
}
such that the EFT expansion converges.

Let us now map this power-counting scheme onto the SMEFT.\footnote{Alternative power counting schemes leading to sizeable self-coupling modifications were, for instance, considered in \cite{Azatov:2015oxa, DiVita:2017eyz}.
We also note that the argument we present in the following does not apply, if the electroweak symmetry is not linearly realised in the EFT.}
At dimension-$6$ and tree-level, we may only have $c_6$.
To generate the other Wilson coefficients at dimension-$6$ requires absorbing two coupling factors with an $\hbar$, and thus a loop suppression.
Since fields ---unlike derivatives--- may also absorb coupling factors, important contributions could also arise at tree-level and dimension-$8$, through operators involving two extra Higgses.
Should such a UV completion exist and be generic, we would expect the following pattern of Wilson coefficients:
\beq
c_{6} \sim \kappa ~~,\qquad c_{H,R,T} \sim \frac{\kappa}{16 \pi^2} ~~,\qquad c_{H_8,R_8,T_8} \sim \kappa ~~,
\eeq
where we have defined the dimension-$8$ operators containing two extra Higgs fields as $\mathcal{O}_{H_8,R_8,T_8} \equiv |H|^2\: \mathcal{O}_{H,R,T}/M^2$.

In terms of observable corrections to Higgs couplings, we then expect the following pattern:
\beq
\delta_{h^3} \sim \kappa \frac{v^4}{M^2 m_h^2} ~~,\qquad \delta_{VV} \sim \kappa \frac{v^2}{M^2} \max \left[ \frac{1}{16 \pi^2}, \frac{v^2}{M^2} \right]  ~~,
\eeq
where $v^2\equiv (\sqrt{2}G_F)^{-1} \approx ( 246 \text{ GeV})^2$.

The generic expectations for contributions to the $\widehat{T}$ parameter, which are
\beq
\Delta \widehat{T} \sim \kappa \frac{v^2}{M^2} \max \left[ \frac{1}{16 \pi^2}, \frac{v^2}{M^2} \right] ~~,
\eeq
may be further suppressed if the UV interactions leading to $\kappa$ respect custodial symmetry.
More on this later.

Based on these arguments, we may sketch an expected \emph{upper bound} on the ratio of Higgs coupling deviations in a generic UV completion
\beq
\left| \frac{\delta_{h^3}}{\delta_{VV}}  \right| \lesssim \min \left[ \left(\frac{4 \pi v}{m_h} \right)^2, \left(\frac{M}{m_h} \right)^2 \right] ~~,
\label{eq:scaling}
\eeq
which reaches $(4\pi v/m_h)^2\approx 600$ for $M\gtrsim 4\pi v\approx 3\,$TeV.
Note that this is a bound and not an expectation since many UV completions, including known models, will not come anywhere close to saturating it without fine-tuning in the UV.
On the other hand, although large ratios are not expected in general, we see that there is in principle ample room for Higgs self-coupling modifications to be significantly greater in magnitude than the modifications of single-Higgs couplings.
In \autoref{sec:tree}, we present an explicit example of a generic UV completion where the bound in \autoref{eq:scaling} is saturated.

\subsection{Higher-dimension potential terms}
\label{sec:eft-expansion}
The previous analysis should hold true for generic UV completions, however there is an additional caveat to this logic that will prove relevant in this work.

In an EFT, there are constraints on the relative magnitude of subsequent terms in the momentum expansion of certain scattering amplitudes.
For instance, UV contributions to identical operator two-point functions or four-point forward scattering amplitudes, captured in a Taylor series as $\sum_j c_j (p^2/M^2)^j$, are known to form a convergent series where $|c_{j+1}|\leq |c_{j}|$, at least from dimension-$8$ onwards \cite{Englert:2019zmt,Bellazzini:2019bzh}.
This means that for a given $c_j$, even if the higher-order terms are unknown, their impact on scattering amplitudes may be bounded from above.

On the other hand, for the Higgs potential, we are interested in the field expansion and not in the momentum expansion.
To our knowledge, in this case, no such convergence condition exists.
Higher-dimension operators may therefore, in principle, contribute to $\delta_{h^3}$ comparably to $\mathcal{O}_6$, or even more significantly.
This could occur even in scenarios which are accurately described by an EFT where the electroweak symmetry is linearly realised.

This caveat raises a potential loophole to general considerations of Higgs coupling deviations based solely on the $\mathcal{O}_6$ operator.
Such an analysis could be overly na\"ive in the sense that it may
{\it a)} underestimate the magnitude of $\delta_{h^3}$ by missing important corrections of higher dimensions,
{\it b)} overestimate vacuum stability constraints on $\delta_{h^3}$ by underestimating the size of stabilising higher-order terms, or, relatedly,
{\it c)} overestimate unitarity constraints on $\delta_{h^3}$ by underestimating unitarity-preserving cancellations from operators of different dimensions that may be present in $n$-point scattering amplitudes as a result of UV symmetries.
The explicit example models discussed in \hyperref[sec:tree]{sections \ref*{sec:tree}} and \ref{sec:pNGB} expose these three caveats.

\subsection{Phenomenological implications}

There are important implications of \autoref{eq:scaling} for Higgs phenomenology.
The foremost arises when we consider what magnitude of self-coupling correction could remain undetected at present.
Consider the current status of LHC measurements, which constrain $|\delta_{VV}| \lesssim 0.07$ at the two-sigma level~\cite{CMS-PAS-HIG-19-005, ATLAS-CONF-2021-053}.
\hyperref[eq:scaling]{Equation~(\ref*{eq:scaling})} thus leaves room in generic UV completions for $|\delta_{h^3}| \lesssim 40$.
On the other hand, present experimental constraints on the Higgs self-coupling are at the level of $-2 <\delta_{h^3} < 5.6$ \cite{CMS-PAS-HIG-19-005, ATLAS-CONF-2021-052}.
Thus, at the moment, self-coupling analyses are actually competitive with single-coupling measurements in probing the parameter space of generic UV completions of the SM.
This will remain true during the rest of the LHC programme.
The HL-LHC is expected to achieve a two-sigma sensitivity%
\footnote{It is customary to quote one-sigma sensitivities for future collider prospects.
To ease comparisons with existing constraints, here and in the following we apply a naive factor of two to approximate two-sigma sensitivities.
For precise measurements, non-Gaussianities can be expected to be moderate.}
of $2.6\%$ on $\delta_{VV}$~\cite{deBlas:2019rxi}, allowing for $|\delta_{h^3}|\lesssim 15$ in generic UV completions, while a $100\%$ two-sigma precision can be expected from Higgs pair production measurements.

Looking further towards the future, if all single-Higgs measurements were consistent with the SM after FCC-ee operation, then one would have $|\delta_{ZZ}| \lesssim 0.34\%$ at the two-sigma level~\cite{deBlas:2019rxi}.
This would imply a generic theoretical upper bound of $|\delta_{h^3}| \lesssim 200\%$, which  indirect constraints from the same machine would already surpass with a $48\%$ sensitivity~\cite{McCullough:2013rea,DiVita:2017vrr,deBlas:2019rxi}.
FCC-hh would moreover probe Higgs self-coupling deviations down to the $10\%$ level at the two-sigma level~\cite{deBlas:2019rxi}.
So, even if Higgs measurements at FCC-ee were SM-like, there would still remain the possibility of finding deviations in Higgs self-coupling measurements at FCC-hh, generated by generic UV completions.

A high-energy lepton collider such as CLIC could achieve $|\delta_{ZZ}|\lesssim 0.78\%$~\cite{deBlas:2019rxi}, leaving open the possibility of having $|\delta_{h^3}| \lesssim 480\%$.
Di-Higgs production with a two-sigma reach of $22\%$ on the trilinear self-coupling~\cite{deBlas:2019rxi} would then cover a sizeable amount of untouched generic parameter space.
The same conclusion would also hold at higher-energy muon colliders, although the associated prospects are more speculative.
With 10~ab$^{-1}$ at 10\,TeV, one could indeed expect single-Higgs coupling precisions in the same ballpark as that of the FCC-ee (0.14\%) and a self-coupling determination with a reach similar to that of the FCC-hh (11\%)~\cite{Han:2020pif}.

In UV completions without custodial protection, the scaling of \autoref{eq:scaling} would also apply to the $|\delta_{h^3}/\Delta \widehat{T}|$ ratio.
A $Z$-pole run at the FCC-ee would improve the current two-sigma constraint on $|\Delta \widehat{T}|$ by an order of magnitude, down to about $10^{-4}$~\cite{deBlas:2019rxi}.
A theoretical upper bound of $|\delta_{h^3}|\lesssim 6\%$ would then arise for generic UV completions.
A one-loop suppression of dimension-6 contributions to custodial symmetry violation is thus insufficient to allow for any testable modification of the Higgs self-coupling in non-custodial UV completions, showing the power of precision $Z$-pole physics.
On the other hand, custodially symmetric UV completions would still allow for sizeable self-coupling modifications, as we discuss in \autoref{sec:custquad}.

Finally, note that the hierarchy in \autoref{eq:scaling} allows self-coupling modifications entering in single-Higgs production and decay processes at one loop factor greater than single-coupling modifications to have an impact on the final observable that is comparable in magnitude.
Thus, for an agnostic analysis of coupling deviations, it is advisable to include higher-order self-coupling effects.
Relatedly, due to the high coupling dimension of $c_6$, such contributions are guaranteed to be finite in single-Higgs production and decay processes, in contrast with the rest of the SMEFT landscape where the inclusion of dimension-6 operators at one-loop will generically lead to a logarithmically divergent contribution.\footnote{This is intimately related to the fact that $\mathcal{O}_6$ renormalises only itself at one loop, as demonstrated in \cite{Jenkins:2013zja,Jenkins:2013wua,Alonso:2013hga}.
Interestingly, the non-renormalisation theorem of \cite{Bern:2019wie} goes further, showing that the renormalisation of operators of the $|H|^4D^2$ class by $|H|^6$ vanishes at two loops.}
These two aspects are fortuitously linked, in that the coupling dimension of $c_6$ allows for a large hierarchy in coupling deviations, but it also permits the higher-loop contributions of self-coupling deviations to physical processes to be finite, lending support to efforts to probe the Higgs self-coupling in this manner \cite{McCullough:2013rea,Gorbahn:2016uoy,Degrassi:2016wml,Bizon:2016wgr, Degrassi:2017ucl, Kribs:2017znd, DiVita:2017eyz,Maltoni:2017ims,DiVita:2017vrr,Maltoni:2018ttu,deBlas:2019rxi,Gorbahn:2019lwq, Degrassi:2019yix, Degrassi:2021uik, Haisch:2021hvy}.

With all of this in mind, let us now explore some generic UV completions which can saturate the bound of \autoref{eq:scaling}.

\section{\texorpdfstring{$|\delta_{h^3}| \gg |\delta_{VV}|$}{delta h3 >> delta VV} at tree level: Custodial weak quadruplet}
\label{sec:tree}
As a first class of generic models realising $|\delta_{h^3}| \gg |\delta_{VV}|$, we consider simple renormalisable scenarios which involve heavy scalar SU(2)$_L$ quadruplets with hypercharge $Y = 1/2$ or $3/2$.
Assuming that these new states are sufficiently heavy for an EFT description to be accurate at low energies, $\mathcal{O}_6$ is the only operator generated at dimension-$6$ and tree level~\cite{deBlas:2014mba,Henning:2014wua,Logan:2015xpa,Jiang:2016czg,Dawson:2017vgm,Corbett:2017ieo,Chala:2018ari,Murphy:2020rsh}.
We proceed by studying in more details the EFT they give rise to.

\subsection{Integrating out weak quadruplets}

A weak quadruplet can be described by the symmetric three-index tensor representation of SU(2)$_L$, $\Phi_{ijk}$.
The four complex canonically normalised propagating degrees of freedom $\omega_{1\, \ldots\, 4}$ are embedded as
\begin{equation}
\begin{aligned}
\Phi_{111} & = \omega_1 ~~, \\
\Phi_{112}  =  \Phi_{211} & =  \Phi_{121}= \omega_2 / \sqrt{3} ~~, \\
\Phi_{122}  =  \Phi_{221} & = \Phi_{212}= \omega_3 / \sqrt{3} ~~,\\
\Phi_{222} & = \omega_4 ~~.
\end{aligned}
\end{equation}
For convenience, we work with the three-index representation.

\subsubsection*{\texorpdfstring{$Y = 1/2$}{Y = 1/2}}
The relevant Lagrangian for $\Phi \sim \mathbf{4}_{1/2}$ is
\begin{equation} \label{eq:quad_Y12}
\mathcal{L}_\Phi = \mathcal{L}_{\rm kin}  - \lambda_\Phi H^{\ast} H^{\ast} (\epsilon H)  \Phi + \mathrm{h.c.}- V_{\mathcal{Z}_2} \qquad , \qquad (\epsilon \equiv i \sigma^2) ~ ,
\end{equation}
where SU(2)$_L$ indices have been suppressed, while the quadratic Lagrangian reads explicitly $\mathcal{L}_{\rm kin} = \text{Tr} [D^\mu \Phi^{\ast} D_\mu \Phi - M_\Phi^2 \Phi^{\ast} \Phi ]$ where $\text{Tr}[\Phi^{\ast} \Phi] = \Phi^{\ast ijk} \Phi_{ijk}$ and the covariant derivative is a six-index tensor.
Due to the hypercharge assignment, the additional terms in $V_{\mathcal{Z}_2}$ respect a $\mathcal{Z}_2$ symmetry acting on $\Phi$.
Note that certain couplings must be present in $V_{\mathcal{Z}_2}$ to ensure vacuum stability (see below).

Assuming $\lambda_\Phi$ to be real and integrating $\Phi$ out at tree level, we obtain
\beq \label{eq:4_EFT6}
\mathcal{L}^6_{\text{EFT}}  =   \frac{\lambda_\Phi^2}{3 M_\Phi^2}  |H|^6 
\eeq
at dimension-$6$, i.e.\ $c_6 = -\lambda_\Phi^2/3$, and
\begin{equation} \label{eq:4_EFT8}
\mathcal{L}^8_{\text{EFT}} = \frac{\lambda_\Phi^2}{3M_\Phi^4}  \big( c_{R_8}\: |H|^4 |D_\mu H|^2 + c_{H_8} \: |H|^2 \partial_\mu |H|^2 \partial^\mu |H|^2 + c_{T_8}\: |H|^2 |H^\dagger D_\mu H|^2 \big)
\end{equation}
at dimension-$8$, with $c_{R_8} = 7, ~c_{H_8} = 2$, and $c_{T_8} = -6$, where the covariant derivative is the appropriate one for the Higgs doublet.%
\footnote{We comment briefly on the relation of our results to previous work.
We have checked the exact equivalence of \autoref{eq:quad_Y12} with~\cite{deBlas:2014mba}, where the interaction of $\Theta_1^J \sim \mathbf{4}_{1/2}$ ($J = 1, 2, 3, 4$) with the Higgs field is written as $\mathcal{L}_{\Theta_1} = - \lambda_{\Theta_1} (H^\dagger \sigma^a H) C_{a\beta}^I (\epsilon H^\ast)_\beta \epsilon_{IJ} \Theta_1^J + \mathrm{h.c.}$ with appropriately defined $C_{a\beta}^I$ and $\epsilon_{IJ}$, provided one identifies $\omega_J = \Theta_1^J$ and $\lambda_\Phi = - \lambda_{\Theta_1}/\sqrt{2}$. Similarly, the notation of~\cite{deBlas:2014mba} for the $\Theta_3^J \sim \mathbf{4}_{3/2}$ is equivalent to ours for $\lambda_{\widetilde{\Phi}} = \sqrt{3/2}\, \lambda_{\Theta_3}$. 

Furthermore, our $c_6$ result for the $\mathbf{4}_{1/2}$ is a factor of $3$ larger than those of ~\cite{Dawson:2017vgm,Corbett:2017ieo}, whereas the correction to the $\widehat{T}$ parameter in~\cite{Corbett:2017ieo} should have an opposite sign.
For the $\mathbf{4}_{3/2}$, our $c_6$ and $c_{T_8}$ results agree with~\cite{Dawson:2017vgm,Corbett:2017ieo}.
We have checked that the $c_6,c_{R_8},c_{H_8},c_{T_8}$ coefficients that we obtain in both $\mathbf{4}_{1/2}$ and $\mathbf{4}_{3/2}$ cases match those of \cite{Murphy:2020rsh} (arXiv v4).
On the other hand, the custodial sum of $\mathbf{4}_{1/2}$ and $\mathbf{4}_{3/2}$ pieces in \cite{Chala:2018ari} has relative $c_6$ and $c_{R_8},c_{H_8}$ contributions differing from ours by a factor of 2.

Finally, the one-loop results reported in~\autoref{eq:4_EFT6oneloop} for the $\mathbf{4}_{1/2}$ agree with the 23 June 2022 update of the \href{https://github.com/effExTeam/Precision-Observables-and-Higgs-Signals-Effective-passageto-select-BSM}{supplementary material} in~\cite{Anisha:2021hgc} (where the quadruplet is denoted by $\Sigma$).
We thank the authors for correspondence about these results.
}
The leading corrections to Higgs couplings, normalised to their SM values are, from \autoref{eq:4_EFT6} at dimension-$6$,
\begin{equation}
\delta_{h^3} = \frac{c_6}{M_\Phi^2 G_F^2m_h^2}~~,
\end{equation}
and from \autoref{eq:4_EFT8} at dimension-$8$
\begin{align}
\delta_{ZZ} =&\,  \frac{\lambda_\Phi^2}{48 G_F^2 M_\Phi^4}\left(4 c_{R_8} - 4 c_{H_8} + 3 c_{T_8}\right) \,, \nonumber \\
\delta_{WW} =&\,   \frac{\lambda_\Phi^2}{48 G_F^2 M_\Phi^4}\left(4 c_{R_8} - 4 c_{H_8} - c_{T_8}\left(3 +\frac{2\bar{s}_w^2}{1- 2 \bar{s}_w^2}\right) \right) \,,  \\
\delta_{ff}=&\, - \frac{\lambda_\Phi^2}{48 G_F^2 M_\Phi^4}\left(4 c_{H_8} + c_{T_8} \right) \,, \nonumber
\end{align}
where we have used $\{\alpha, m_Z, G_F\}$ as electroweak input parameters and defined
\beq
\bar{s}_w^2 \equiv \tfrac{1}{2} \left( 1 - \sqrt{1 - \tfrac{4\pi \alpha}{\sqrt{2}G_F m_Z^2}}\,\right)~.
\eeq
Note that $c_{T_8} \neq 0$ implies $\delta_{ZZ} \neq \delta_{WW}$, signalling the violation of custodial symmetry.
The latter is most strongly constrained by the electroweak $\widehat{T}$ parameter, which is corrected as
\begin{equation}
\Delta \widehat{T} =  - \frac{c_{T_8} \lambda_\Phi^2}{24 G_F^2 M_\Phi^4}
\end{equation}
at tree level.

\begin{figure}\centering
\fmfframe(3,3)(3,3){\begin{fmfgraph*}(30,15)
\fmfleft{l1,l2}
\fmfright{r1,r2}
\fmf{dashes,tens=2}{l1,v1,l2}
\fmf{dashes,tens=2}{r1,v2,r2}
\fmf{dashes,left,lab=$H$}{v1,v2}
\fmf{vanilla,right,lab=$\Phi$}{v1,v2}
\fmfdot{v1}
\fmfdot{v2}
\end{fmfgraph*}}
\fmfframe(3,3)(3,3){\begin{fmfgraph*}(30,15)
\fmfleft{l1,l2}
\fmfright{r1,r2}
\fmf{dashes,tens=2}{l1,v1,l2}
\fmf{dashes,tens=2}{r1,v2,r2}
\fmf{vanilla,left,lab=$\Phi$}{v1,v2}
\fmf{vanilla,right,lab=$\Phi$}{v1,v2}
\fmfdot{v1}
\fmfdot{v2}
\end{fmfgraph*}}
\fmfframe(3,3)(3,3){\begin{fmfgraph*}(30,15)
\fmfleft{l1,l2}
\fmfright{r1,r2}
\fmf{dashes,tens=2}{l1,v1}
\fmf{dashes,tens=2}{r1,v2}
\fmf{dashes,tens=2}{l2,w1}
\fmf{dashes,tens=2}{r2,w2}
\fmf{dashes,tens=1}{w1,v1}
\fmf{dashes,tens=1}{w2,v2}
\fmf{photon,lab=$B$, lab.side=left}{w1,w2}
\fmf{dashes}{v1,v2}
\fmfdot{v1}
\fmfdot{v2}
\fmffreeze
\fmf{vanilla,lab=$\Phi$, right, lab.side=right}{v1,v2}
\end{fmfgraph*}}
\fmfframe(3,3)(3,3){\begin{fmfgraph*}(30,15)
\fmfleft{l1,l2,l3}
\fmfright{r1,r2,r3}
\fmf{dashes,tens=2}{l1,v1}
\fmf{dashes,tens=2}{r1,v2}
\fmf{dashes,tens=0}{v2,r2}
\fmf{dashes,tens=0}{v1,l2}
\fmf{dashes,tens=2}{l3,w1}
\fmf{dashes,tens=2}{r3,w2}
\fmf{dashes,tens=1}{w1,v1}
\fmf{dashes,tens=1}{w2,v2}
\fmf{photon,lab=$B$, lab.side=left}{w1,w2}
\fmf{vanilla,lab=$\Phi$, lab.side=right}{v1,v2}
\fmfdot{v1}
\fmfdot{v2}
\fmffreeze
\end{fmfgraph*}}
\caption{Illustrative diagrams involved in the loop-level matching of the weak quadruplet models to Higgs operators violating custodial symmetry at dimension-6 and -8.}
\label{fig:one-loop-matching}
\end{figure}
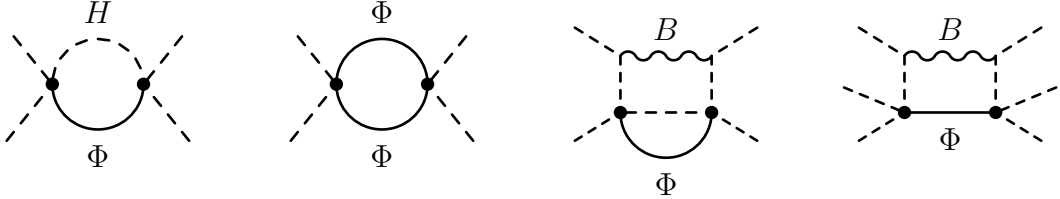

One also expects dimension-$6$ corrections at the one-loop level which, depending on the separation between the electroweak and new physics scales, may be more important than the dimension-$8$ tree-level contributions.
From the $\lambda_\Phi$ interaction at one-loop, we find
\begin{equation} \label{eq:4_EFT6oneloop}
\mathcal{L}^6_{\text{EFT}} = \frac{\lambda_\Phi^2}{12 \pi^2 M_\Phi^2}  \big( c_R^{(1)} \: |H|^2 |D_\mu H|^2 + c_H^{(1)}  \: \partial_\mu |H|^2 \partial^\mu |H|^2 + c_T^{(1)} \: |H^\dagger D_\mu H|^2 \big) ~,
\end{equation}
where $c_R^{(1)} = 5,~ c_H^{(1)} = 1$, and $c_T^{(1)} = -3$. The associated Higgs coupling corrections are 
\begin{align}
\delta_{ZZ} =&\,  \frac{\lambda_\Phi^2}{48 \pi^2 \sqrt{2} G_F M^2_\Phi}\left(2 c_R^{(1)} - 4 c_H^{(1)} + c_T^{(1)} \right) \,, \nonumber \\
\delta_{WW} =&\,   \frac{\lambda_\Phi^2}{48 \pi^2 \sqrt{2} G_F M^2_\Phi}\left(2 c_R^{(1)} - 4 c_H^{(1)} 
 - c_T^{(1)} \left( 3 + \frac{2\bar{s}_w^2}{1-2\bar{s}_w^2}\right)
\right) \,, \\
\delta_{ff}=&\, - \frac{\lambda_\Phi^2}{48 \pi^2 \sqrt{2} G_F M^2_\Phi} \left(4 c_H^{(1)} + c_T^{(1)} \right) \nonumber \,,
\end{align}
and the $\widehat{T}$ parameter is
\begin{equation}
\Delta \widehat{T} =  - \frac{c_{T}^{(1)} \lambda_\Phi^2}{24\pi^2 \sqrt{2} G_F M_\Phi^2}~.
\end{equation}
Since we have not specified $V_{\mathcal{Z}_2}$, we will not calculate the corrections originating from it, noting that they scale in the same way.
Thus, if the quartic couplings in $V_{\mathcal{Z}_2}$ are comparable in magnitude to $\lambda_\Phi$, they may be equally as important (see the first two diagrams in \autoref{fig:one-loop-matching}).
We focus on those generated by $\lambda_\Phi$ since they are an irreducible contribution directly correlated with the Higgs self-coupling correction.

Other operators are also generated at the one-loop level, including those containing two derivatives and two electroweak field-strength tensors, which are related to the $W$ and $Y$ parameters.
In particular, one finds $W = g_2^2 m_W^2 / (96\pi^2 M_{\Phi}^2)$~\cite{Henning:2014wua,Anisha:2021hgc}.
Existing constraints from LEP2 (and from Drell-Yan processes at the LHC, where the high-energy tail lies beyond the EFT validity) are not competitive with direct searches for the quadruplet states, see \autoref{sec:direct-searches}.
This will remain true at future colliders.
In addition, operators involving two electroweak field-strength tensors and four/two Higgs fields are generated at the one-loop/two-loop levels, respectively.
They notably contribute to the Higgs decay to two photons, which is also loop-induced in the SM.
The resulting sensitivity is however not expected to surpass that of other Higgs coupling measurements.

\subsubsection*{\texorpdfstring{$Y = 3/2$}{Y = 3/2}}
For the $\widetilde{\Phi} \sim \mathbf{4}_{3/2}$ quadruplet, the UV Lagrangian reads
\begin{equation}\label{eq:quad_Y32}
\mathcal{L}_{\widetilde{\Phi}} = \mathcal{L}_{\rm kin}  - \frac{\lambda_{\widetilde{\Phi}}}{\sqrt{3}} H^{\ast} H^{\ast} H^{\ast}  \widetilde{\Phi} + \mathrm{h.c.} - V_{\mathcal{Z}_2}~~,
\end{equation}
with the obvious replacements in $\mathcal{L}_{\rm kin}$.
The normalisation of the coupling is chosen such that the resulting tree-level effective theory is analogous to \hyperref[eq:4_EFT6]{eqs.\,\eqref{eq:4_EFT6}} and~\hyperref[eq:4_EFT8]{\eqref{eq:4_EFT8}}, with $\lambda_\Phi\to\lambda_{\widetilde{\Phi}}$ and $M_\Phi \to M_{\widetilde{\Phi}}$, but now $c_{R_8} = 3,~ c_{H_8} = 0\,$, and $c_{T_8} = 6$.
At one loop, we find $c_R^{(1)} = 3,~ c_H^{(1)} = 0\,$, and $c_T^{(1)} = 3$ in the conventions of \autoref{eq:4_EFT6oneloop}.

\begin{figure}[t]\centering
\adjustbox{max width=.95\textwidth}{%
\begin{tabular}{@{}c@{}}%
\includegraphics{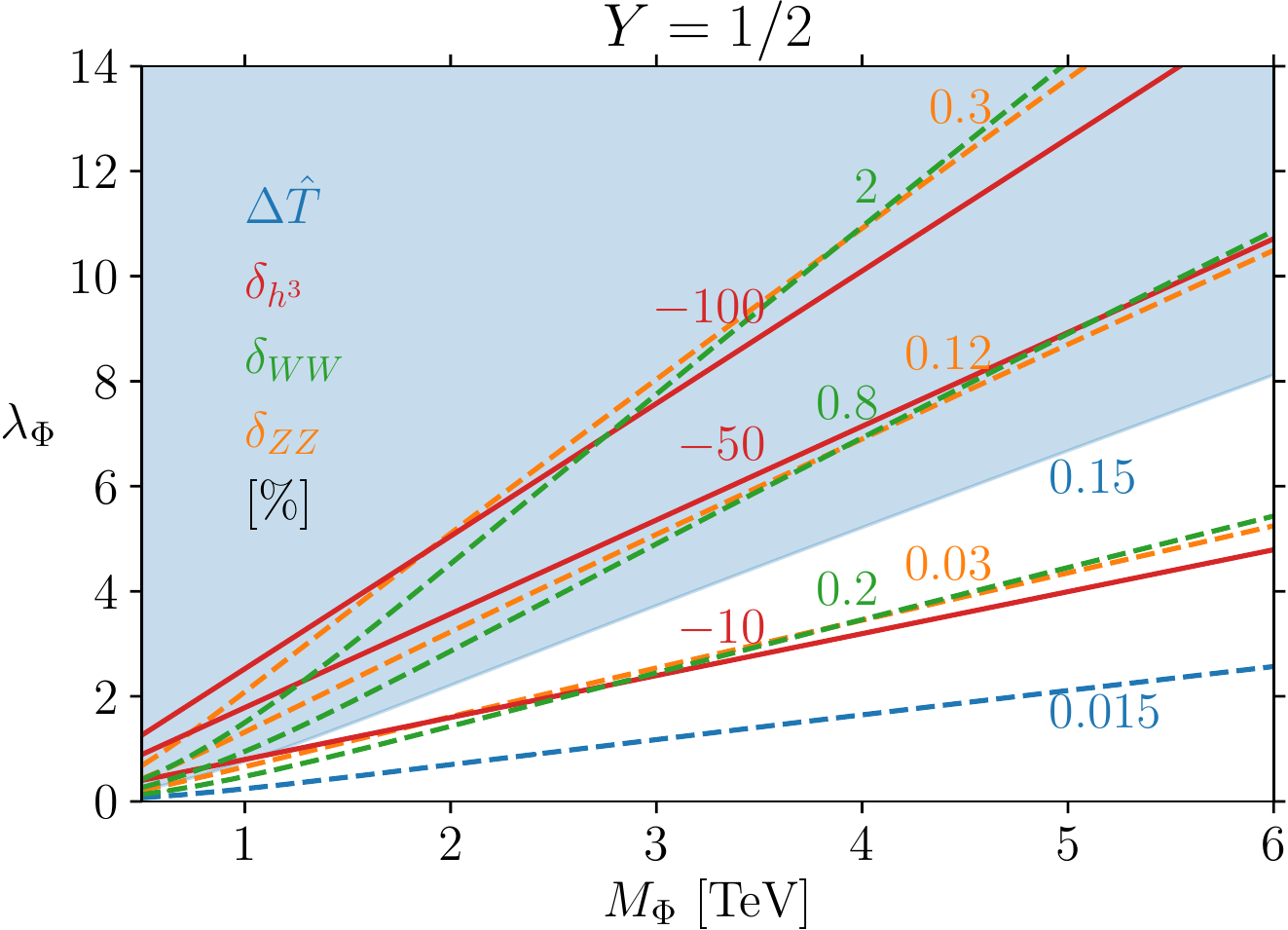}%
\qquad%
\includegraphics{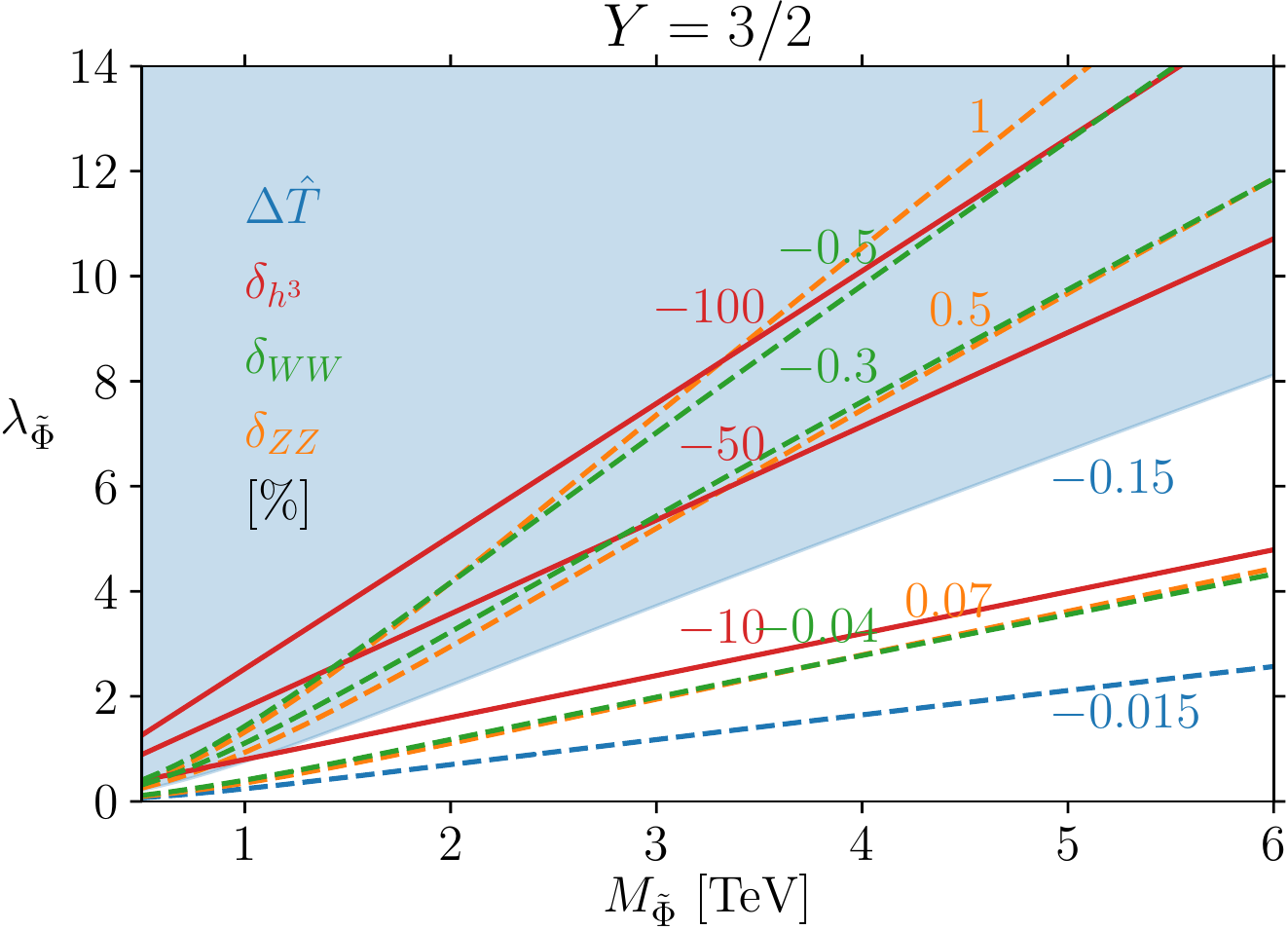}%
\end{tabular}%
}
\caption{Parameter space of the scalar weak quadruplet models with $Y = 1/2$ (left) and $Y = 3/2$ (right).}
\label{fig:custviol}
\end{figure}

\vspace{1mm}
In \autoref{fig:custviol}, we show the Higgs coupling corrections and precision electroweak constraints on these two custodial violating models.
For $M_\Phi/\lambda_\Phi$ of $\mathcal{O}(1)\,\text{TeV}$, self-coupling deviations of tens of percent are compatible with current $|\Delta \widehat{T}|<0.15\%$ constraints~\cite{Workman:2022ynf}\,\footnote{This is an approximate two-sigma bound, which neglects the slight preference of data for positive $\widehat{T}$~\cite{Workman:2022ynf}.
This preference would become strong if we were to include the recent CDF II measurement of the $W$ mass~\cite{CDF:2022hxs}.
Given the unclear current picture of the electroweak fit, we prefer to make a conservative choice.}
and are only accompanied by sub-percent $\delta_{VV}$ corrections.
On the other hand, a future Tera-$Z$ stage at FCC-ee leading to $|\Delta \widehat{T}|<0.015\%$ would exclude the majority of parameter space with significant Higgs self-coupling deviations, showing the powerful synergy between a Tera-$Z$ programme and Higgs physics.

\subsection{The custodial quadruplet}
\label{sec:custquad}
Custodial symmetry is violated in the previous two models, at dimension-$8$ and tree level, as well as at dimension-$6$ and one-loop level.
As discussed in \autoref{sec:intro} and confirmed above, a one-loop suppression of dimension-$6$ effects is insufficient to allow for a modification of the Higgs self-coupling larger than $\mathcal{O}(10) \%$ in custodial violating UV completions.
We thus turn our attention to a custodially symmetric quadruplet model~\cite{Logan:2015xpa,Chala:2018ari}.

The custodial symmetry is $\text{SO}(4) \simeq \text{SU}(2)_L \times \text{SU}(2)_R$, broken in the SM by the Yukawas and the hypercharge gauge coupling.
Both previously considered complex quadruplets naturally arise as a $\mathbf{16}$ irreducible representation of $\text{SO}(4)$, coupled as $\hat{\Phi}^{abc} h_a h_b h_c $ where the four real degrees of freedom of the Higgs are written as a fundamental of $\text{SO}(4)$, $h_a$.
Similarly, in terms of $\text{SU}(2)_L \times \text{SU}(2)_R$ the two quadruplets arise in the decomposition of the $(\mathbf{4},\mathbf{4})$, coupled as $\hat{\Phi}_{ijk}^{\;\;\;\;IJK} \mathcal{H}^{\ast i}_{\;\;\;I} \mathcal{H}^{\ast j}_{\;\;\;J} \mathcal{H}^{\ast k}_{\;\;\;K}$, where we have written the four Higgs degrees of freedom as furnishing a bifundamental $(\mathbf{2},\mathbf{2})$ representation, $\mathcal{H}$.
At the renormalisable level, the potential of the custodial quadruplet model may contain, in addition to $\hat{\Phi}\mathcal{H}^3$, also $\hat{\Phi}^2 \mathcal{H}^2$, $\hat{\Phi}^4$, and $\hat{\Phi}^3 \mathcal{H}$ terms~\cite{Logan:2015xpa} (to preserve invariance under hypercharge, the last operator always involves both $\Phi$ and $\widetilde{\Phi}$).
We do not include these terms in our analysis, but discuss their role in ensuring vacuum stability below.

A possible UV motivation for this setup would be a model where both the Higgs and custodial quadruplet fields emerge among the pNGBs of a new strong sector.
For instance, a spontaneous global $\text{SO}(21)\to \text{SO}(20)$ breaking yields a $\mathbf{20}$ of Goldstone bosons.
These could be split into $(\mathbf{4},\mathbf{1})+(\mathbf{1},\mathbf{16})$ of the block-diagonal $\text{SO}(4)\times\text{SO}(16)$ subgroup, which are also $(\mathbf{4},\mathbf{1})+(\mathbf{1},\mathbf{16})$ of a suitably-defined $\text{SO}(4)\times\text{SO}(4)$ subgroup.
Under the diagonal $\text{SO}(4)$ of the latter, the pNGBs would thus transform as a $\mathbf{4}$ and a $\mathbf{16}$.
In the following, we will however remain agnostic about a deeper UV origin of the custodial quadruplet.

Integrating out the heavy $\hat{\Phi}$, we expect to generate $\mathcal{O}_6$ together with very suppressed custodial violation.
We may show this by recycling the previous results, since the decomposition of the $(\mathbf{4},\mathbf{4})$ of $\text{SU}(2)_L \times \text{SU}(2)_R$ (or equivalently, the $\mathbf{16}$ of $\text{SO}(4)$) under the SM electroweak gauge group leads to the two complex scalar fields 
\beq
(\mathbf{4},\mathbf{4}) \;\;\to\;\; \mathbf{4}_{1/2} + \mathbf{4}_{3/2} ~~,
\eeq
under $\text{SU}(2)_L \times \text{U}(1)_Y$.
The explicit embedding is provided in \autoref{sec:custodial}.
Due to custodial symmetry, the two scalar multiplets couple to the Higgs as
\bea
\mathcal{L}_{\text{SO}(4)}  & = &   - \lambda \Big( H^\ast H^\ast (\epsilon H) \Phi  + \frac{1}{\sqrt{3}} H^\ast H^\ast H^\ast  \widetilde{\Phi}\Big) + \mathrm{h.c.}
\label{eq:custodialcoup}
\eea
and have equal mass.
In other words, the custodial model gives rise to the sum of \hyperref[eq:quad_Y12]{eqs.~(\ref{eq:quad_Y12})} and \hyperref[eq:quad_Y32]{\eqref{eq:quad_Y32}} with $\lambda_\Phi = \lambda_{\widetilde{\Phi}} = \lambda$ and $M_\Phi = M_{\widetilde{\Phi}} = M$, thus generating at tree level the effective operators
\beq \label{eq:4_EFT6cust}
\mathcal{L}^6_{\text{EFT}}  =   \frac{2 \lambda^2}{3 M^2}  |H|^6  ~~,
\eeq
i.e.\ $c_6 = -2\lambda^2/3$ at dimension-$6$, and
\beq \label{eq:4_EFT8cust}
\mathcal{L}^8_{\text{EFT}}  =   \frac{2 \lambda^2}{3 M^4}  \big( 5 |H|^4 |D_\mu H|^2 + |H|^2 \partial_\mu |H|^2 \partial^\mu |H|^2  \big) ~~,
\eeq
i.e.\ $c_{R_8} = 10, ~c_{H_8} = 2$, and $c_{T_8} = 0$ at dimension-$8$.
As anticipated, this is custodially symmetric.
Furthermore, at one-loop and dimension-$6$ we find
\bea
\mathcal{L}^6_{\text{EFT}} =  \frac{\lambda^2}{12 \pi^2 M^2}  \big( 8  |H|^2 |D_\mu H|^2 +  \partial_\mu |H|^2 \partial^\mu |H|^2  \big) ~,
\eea
i.e.\  $c_R^{(1)} = 8,~ c_H^{(1)} = 1$, and $c_T^{(1)} = 0$, which also respects custodial symmetry.
The custodial symmetry persists at one-loop and dimension-$6$ due to the power counting, since Wilson coefficients can involve only $\lambda^2$ and no other coupling at this order.
As the scalar interactions are custodially symmetric, they alone cannot generate $\mathcal{O}_T$. 
Thus we expect custodial symmetry violation to occur only at two loops and dimension-$6$ or one loop and dimension-$8$ (see the last two diagrams in \autoref{fig:one-loop-matching}).

The magnitude of these higher-loop effects can be estimated by examining the known renormalisation-group mixing into custodial symmetry violating operators.
From Appendix~C.3 of \cite{Alonso:2013hga}, the mixing of $c_{H\square}$ into $c_{HD}$, defined as the coefficients of the Warsaw basis dimension-6 operators $Q_{H\square} = - \mathcal{O}_H$ and $Q_{HD} = \mathcal{O}_T$, is proportional to $5g_1^2/12\pi^2$.
The anomalous dimension matrix between dimension-8 bosonic operators has been studied in \cite{AccettulliHuber:2021uoa, DasBakshi:2022mwk}.
In the latter reference, $c_{\phi^6}^{(2)}$ contains a custodial violating component (see the definition in Table~1 of \cite{Chala:2021cgt}).
According to the associated \href{https://github.com/SMEFT-Dimension8-RGEs/Notebooks}{ancillary material}, it receives a mixing from $c_{\phi^6}^{(1)}$ proportional to $5 g_1^2/24\pi^2$.
The mixing of two dimension-6 operators into a dimension-8 one, computed in \cite{Chala:2021pll}, would be subleading in our case.
For $M$ in the few-TeV range, $c_{H_8}$ and $c_H^{(1)}$ of $\mathcal{O}(1)$ at that scale would therefore induce a custodial violation through $c_{T_8}$ and $c_T^{(1)}$ of $\mathcal{O}(1)\%$ at the $Z$-pole.
Contributions to the $\widehat{T}$ parameter are therefore expected to be further suppressed by about two orders of magnitude compared to the custodial violating cases.
At this level of suppression, not even the electroweak precision measurements of the FCC-ee would be competitive with Higgs coupling determinations at the LHC.

\subsubsection*{Vacuum stability}
\addcontentsline{toc}{subsubsection}{\texorpdfstring{$\quad\;\;$}{}Vacuum stability}

Vacuum stability requirements become relevant in regions with significant corrections to the Higgs self-coupling (see e.g.~\cite{Degrassi:2016wml,DiLuzio:2017tfn,Falkowski:2019tft}).
Let us follow the approach of~\cite{Falkowski:2019tft} and consider, in addition to $\mathcal{O}_6$, a dimension-8 contribution to the Higgs potential:
\begin{equation}
- \mathcal{L}_{\rm BSM} \; =\; \frac{c_6}{M^2} |H|^6 + \frac{c_8}{M^4} |H|^8~~.
\label{eq:c6c8}
\end{equation}
This is sufficient for ensuring vacuum stability at small field values.
The situation at large field values could depend on even higher-dimension operators, or need to be studied in the full UV model.

Vacuum stability at small field values qualitatively demands that a large $h^3$ coupling be compensated by an even larger $h^4$ one  (scaling as the square of $\delta_{h^3}$, see details in \autoref{sec:vacsta}).
In turn, this requires that $\delta_{h^3}$ receives a minimal contribution from the dimension-8 operator $|H|^8$.
Decomposing $\delta_{h^3}$ into pieces arising respectively from $|H|^6$ and $|H|^8$,
\begin{equation}
\delta_{h^3} = \delta_{h^3}^{(6)} + \delta_{h^3}^{(8)}\;,\qquad\quad \delta_{h^3}^{(6)} = \frac{2 c_6 v^4}{M^2m_h^2}~,\quad \delta_{h^3}^{(8)} = \frac{4 c_8 v^6}{M^4m_h^2}~,
\label{eq:dh3-decomposition}
\end{equation}
one obtains for $c_6 < 0$, as in the custodial quadruplet model, that
\begin{equation}
\delta_{h^3}^{(8)} \ge -\delta_{h^3}^{(6)} + 1 - \sqrt{1-2\delta_{h^3}^{(6)}}
\qquad\text{for }\delta_{h^3}^{(6)}<0 \text{ and } \delta_{h^3}^{(8)}<-\delta_{h^3}^{(6)}~~.
\label{eq:limit-on-negative-r}
\end{equation}
Hence, beyond $\delta_{h^3}^{(6)}\le-400\%$, the relative dimension-8 contribution to the self-coupling starts becoming significant: $\big| \delta_{h^3}^{(8)}/\delta_{h^3}^{(6)} \big|_{\rm min} \geq 1/2$.
An assessment of the magnitude of $\delta_{h^3}$ based only on the dimension-6 contribution then becomes inaccurate.
More generally, the EFT expansion could also start breaking down at this point, since contributions from different orders would become of similar magnitude.

In the custodial quadruplet model, ignoring quartic potential terms with more than one power of $\hat{\Phi}$ (such as $ |\hat{\Phi}|^2 |H|^2$, $\hat{\Phi}^3 H$, or $|\hat{\Phi}|^4$) is therefore no longer justified when the trilinear Higgs self-coupling computed from $|H|^6$ reaches large negative values.
In this region, including the minimal dimension-8 contributions required for vacuum stability would for example take the dimension-6 estimate of $\delta_{h^3}^{(6)}=-400\%$ to \mbox{$\delta_{h^3}=1-(1-2\,\delta_{h^3}^{(6)})^{1/2}=-200\%$}.
Additionally, the sizeable $ |\hat{\Phi}|^2 |H|^2$ quartic needed to generate the necessary $|H|^8$ operator would give rise to significant contributions to single-Higgs couplings.
For $\delta_{h^3}^{(6)}\lesssim-400\%$, our analysis based on the sole $\hat{\Phi} H^3$ quartic potential term is thus no longer reliable.
Based on estimates within a toy version of the custodial model, we however expect the $\delta_{h^3}/\delta_{VV}$ ratio to be more robust against these additional quartic corrections than the individual coupling modifications.

\subsubsection*{Perturbative unitarity}
\addcontentsline{toc}{subsubsection}{\texorpdfstring{$\quad\;\;$}{}Perturbative unitarity}
As discussed in~\cite{Falkowski:2019tft}, the large $|H|^8$ contribution which ensures vacuum stability when $\delta_{h^3}$ becomes sizeable is constrained by perturbative unitarity.
Vacuum stability for instance requires
\begin{equation}
-\delta_{h^3}
	\le \frac{2\sqrt{2c_8}\,v^3}{M^2 m_h}
	\approx \frac{\sqrt{c_8}}{(4\pi)^3}
		\left(\frac{26\,\text{TeV}}{M}\right)^2
\label{eq:dh3-intervals-simple}
\end{equation}
in the region where $\delta_{h^3} < 0$ and $c_8 > 0$.
Recalling that $c_8$ carries six powers of coupling dimension and using the traditional value of $4\pi$ as the strong coupling limit, $M$ must therefore be pushed to more than $20\,$TeV to obtain constraints of order one on $\delta_{h^3}$.
For new physics in the TeV range and self-coupling deviation of a few, vacuum stability does therefore not seem to push us out of the perturbative regime.

A stronger requirement derives from the model-independent analysis of tree-level unitarity in multi-boson scattering~\cite{Chang:2019vez}, leading to
\beq \label{eq:Chang_Luty}
|\delta_{h^3}| \lesssim \left(\frac{13.4 \text{ TeV}}{E_{\text{max}}}\right)^2~~,
\eeq
where $E_{\text{max}}$ is the energy scale at which perturbativity is lost.
It indicates that self-coupling deviations of order one require new physics to arise around 10\,TeV.
Comparing \autoref{eq:Chang_Luty} to the dimension-6 relation,
\begin{equation}
\delta_{h^3}^{(6)}
	= \frac{2c_6 v^4}{M^2 m_h^2}
	\approx \frac{c_6}{(4.4)^4} \left(\frac{13.4\,\text{TeV}}{M}\right)^2~~,
\end{equation}
indicates that strong coupling could be reached towards $4.4$ instead of $4\pi$.
In the custodial quadruplet model where $c_6 = -2\lambda^2/3$, this corresponds approximately to $\lambda\approx 8\pi$ which is parametrically obtained by demanding $|\mathrm{Re}\,a_0 | < 1/2$ on the $l=0$ partial wave of the $H H \to \hat{\Phi} H$ scattering.

\subsubsection*{Direct searches}
\begin{figure}[t]\centering
\adjustbox{max width=.95\textwidth}{%
\begin{tabular}{@{}c@{}}%
\includegraphics{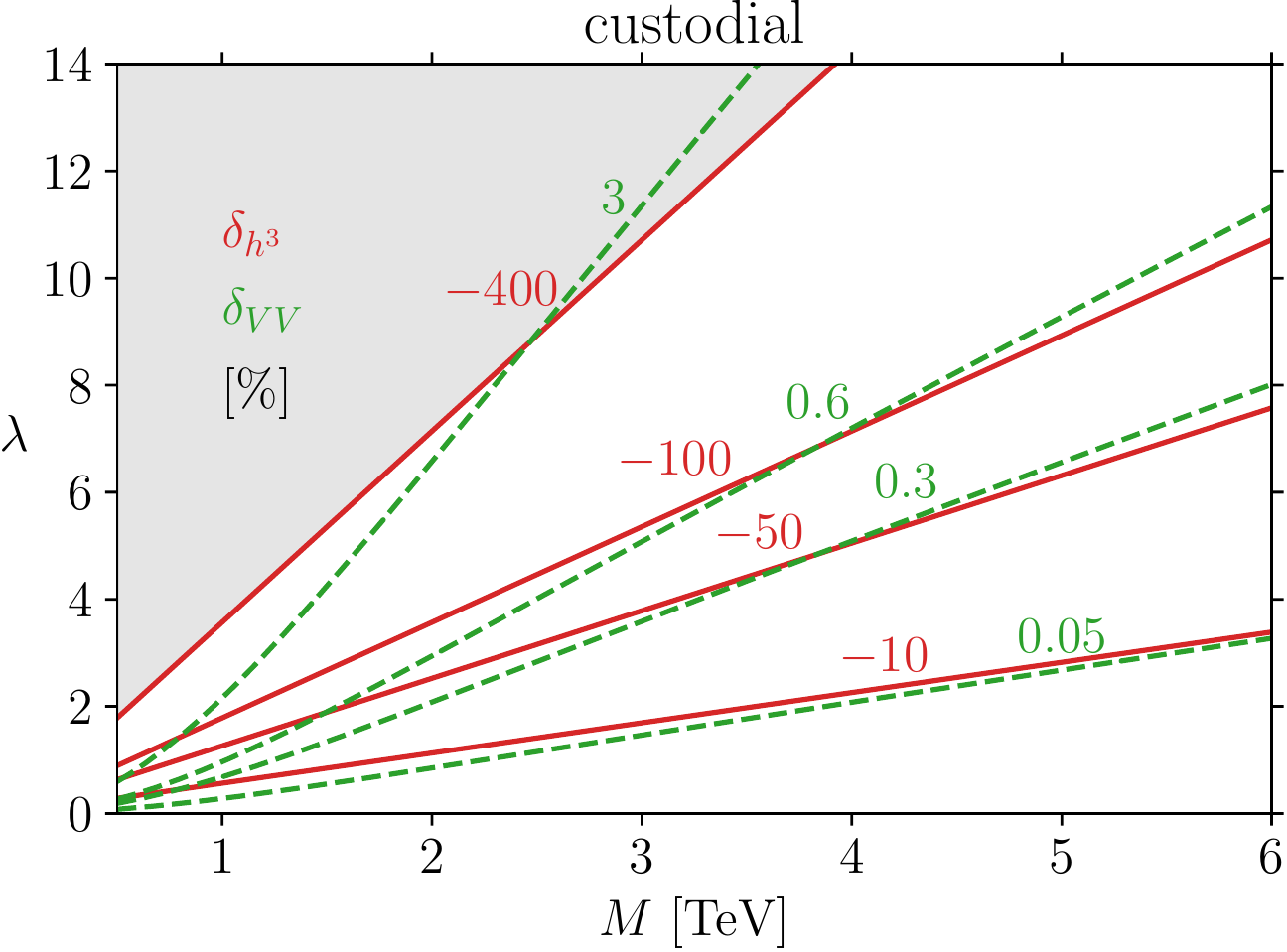}%
\qquad%
\includegraphics{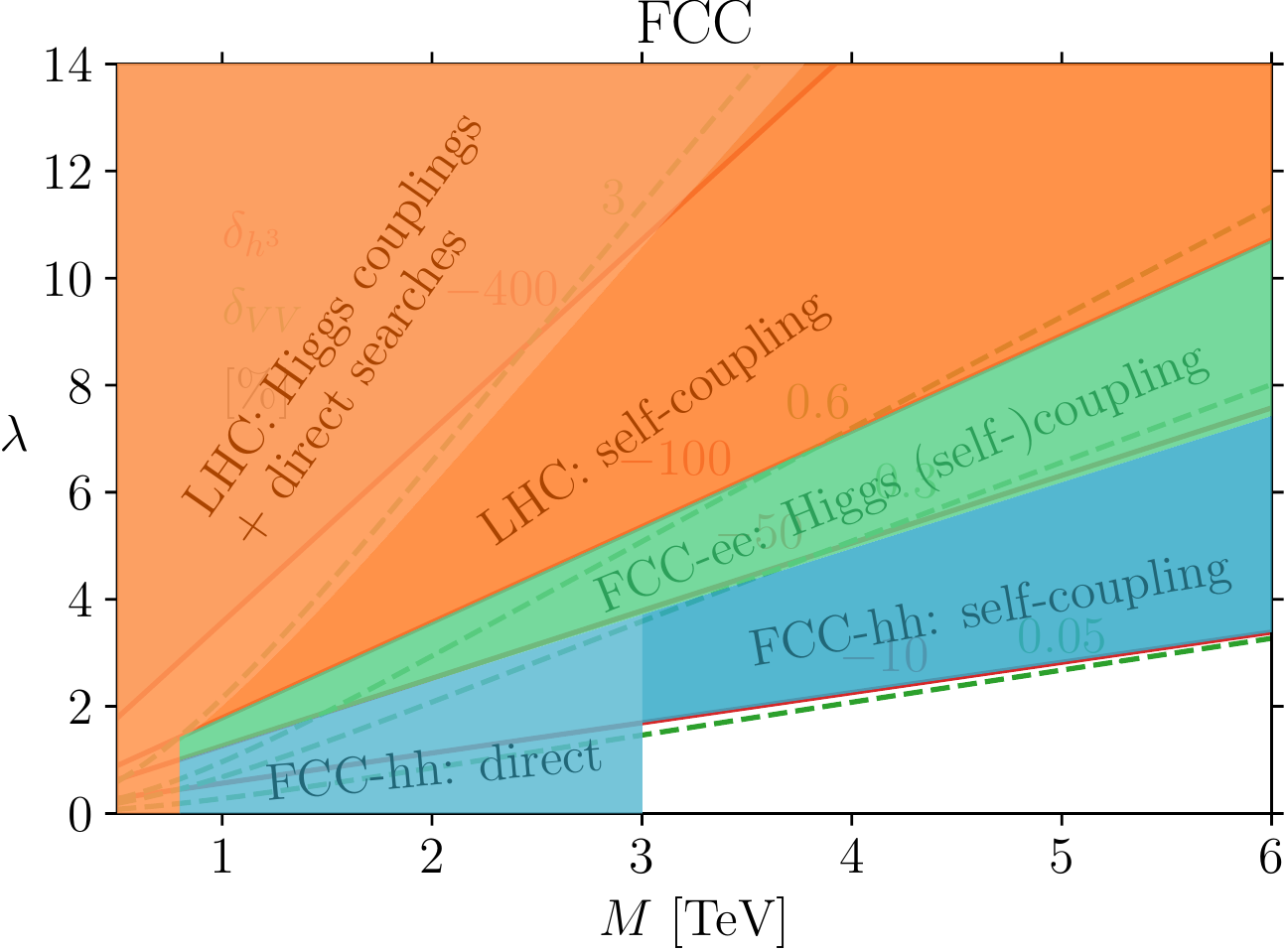}\vspace{5mm}
\\%
\includegraphics{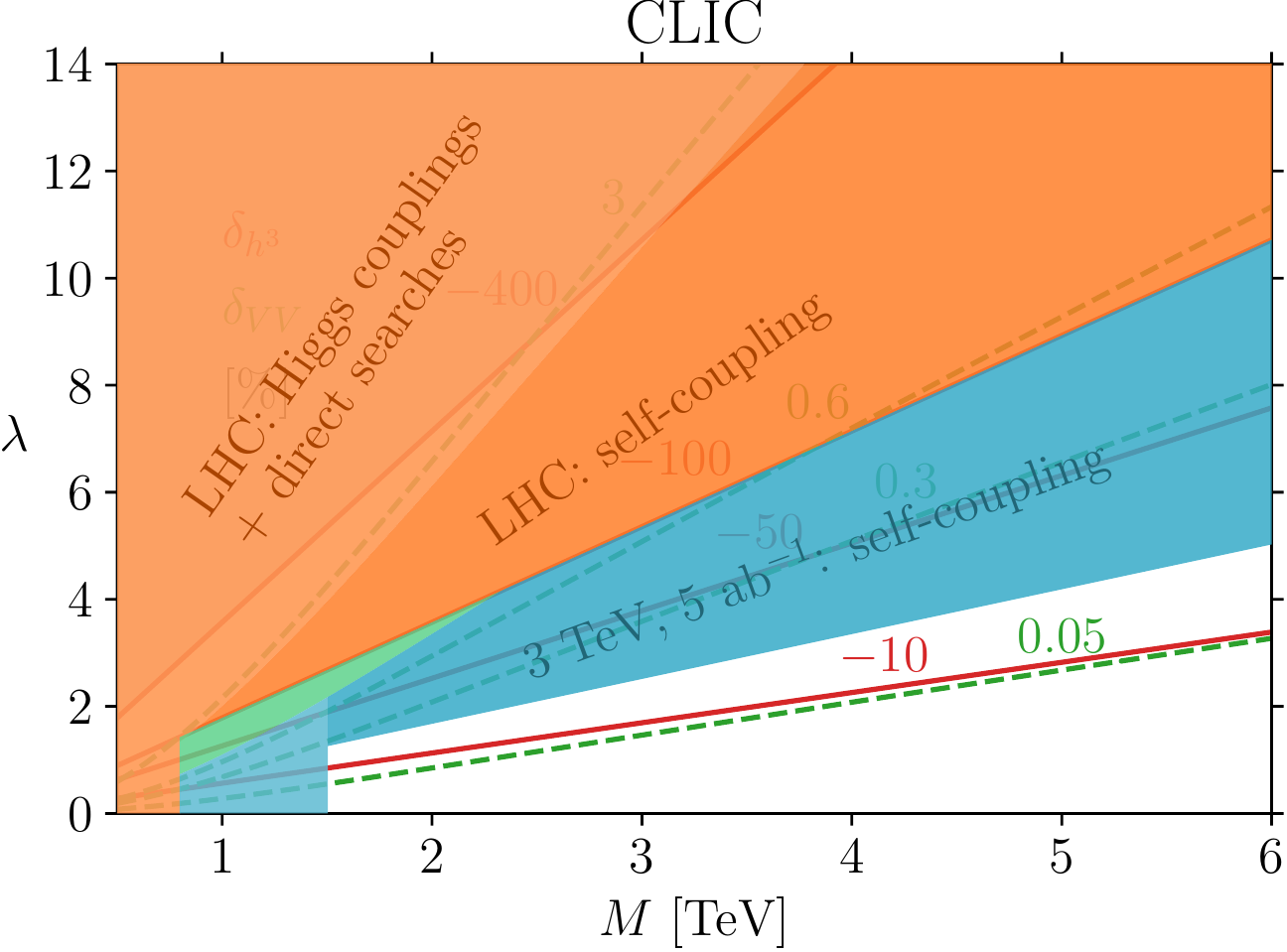}%
\qquad%
\includegraphics{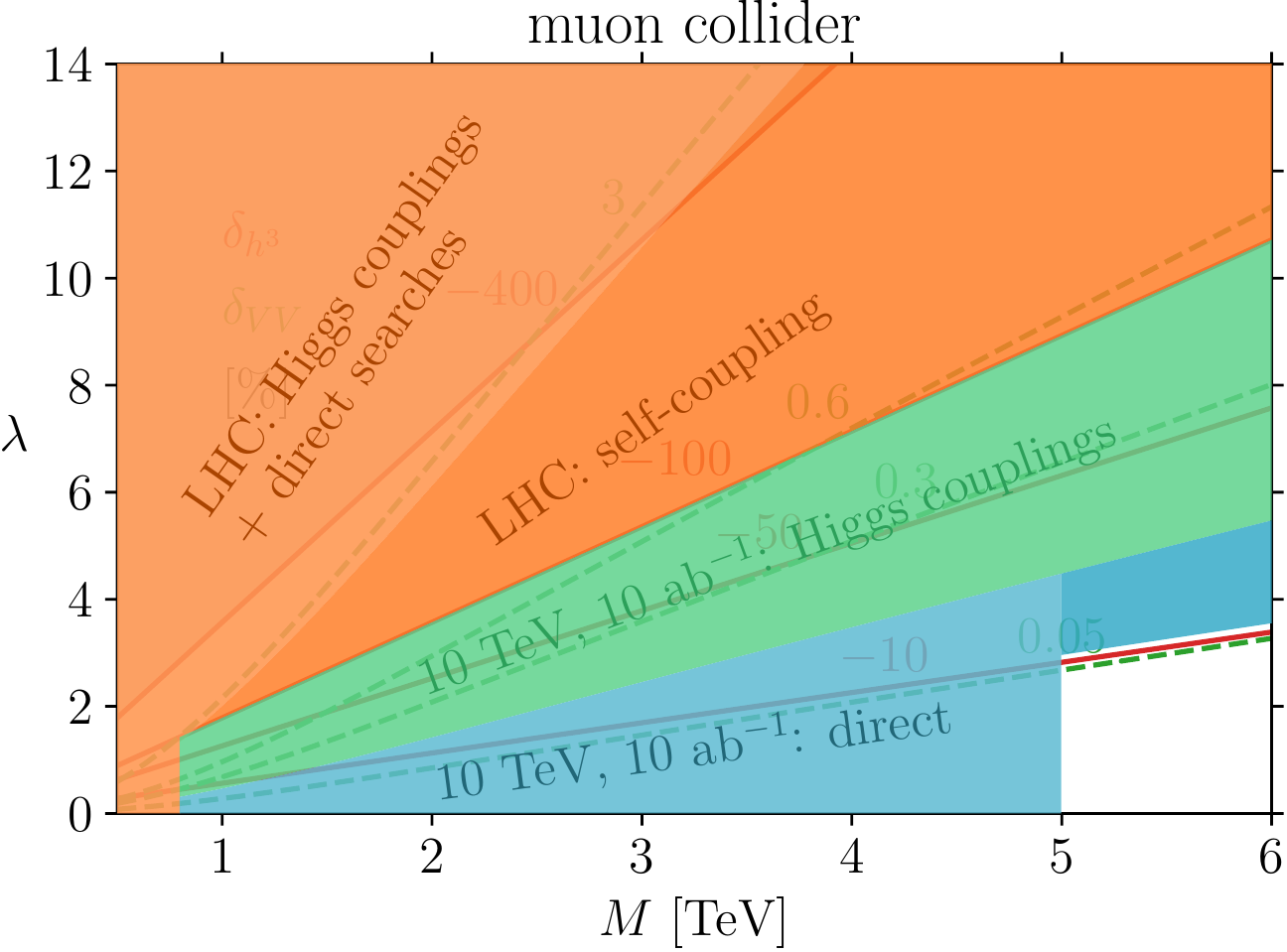}%
\end{tabular}%
}%
\caption{Parameter space of the custodial quadruplet model.
{\it Top left:} Single-Higgs coupling and self-coupling deviations.
In the grey region, vacuum stability requires sizeable contributions from potential terms not included in our analysis.
They would affect both $\delta_{h^3}$ and $\delta_{VV}$ coupling modifications by order-one factors, although their ratio may be relatively stable.
{\it Top right:} Regions probed, at the two-sigma level, by different types of future measurements at the LHC and FCC.
Some prospects are only qualitatively estimated rather than robustly established.
{\it Bottom left:} Same, but assuming a 3\,TeV lepton collider as future project.
{\it Bottom right:} Same, but assuming a 10\,TeV muon collider as future project.
}
\label{fig:quadruplets}
\end{figure}

\addcontentsline{toc}{subsubsection}{\texorpdfstring{$\quad\;\;$}{}Direct searches}
\label{sec:direct-searches}
Beside Higgs (self-)coupling measurements, direct searches for the various components of the custodial quadruplet provide complementary probes of the $(M,\lambda)$ parameter space.
The earliest probes in different regions are summarised in \autoref{fig:quadruplets} for the LHC, FCC, and lepton colliders with either 3 or 10\,TeV centre-of-mass energies.

At the LHC, pair production of electroweak states has a reach in mass of a few hundreds of GeV.
An ATLAS search exploiting the full run-2 dataset for instance sets a two-sigma bound at 350\,GeV on the mass of the doubly charged scalar of the Georgi-Machacek model~\cite{ATLAS:2021jol}.
The HL-LHC two-sigma reach estimated in \cite{Chala:2018ari} for the custodial quadruplet is just below 600\,GeV.
At FCC-hh, masses up to about 3\,TeV can likely be probed.
In the left-right symmetric model, the pair production of doubly charged scalars decaying to lepton pairs was for instance estimated to have a three-sigma reach of 800\,GeV at the HL-LHC and between 2.5 and 6\,TeV at the FCC-hh~\cite{Dev:2016dja}.
Instead of charged leptons, the doubly charged components of the custodial quadruplet would mostly decay into pairs of gauge bosons.

The single production of doubly and singly charged states through vector boson fusion was searched for by CMS, in the $W^\pm W^\pm$ and $W^\pm Z$ channels, using the run-2 double-differential $m_{jj},m_T^{VV}$ distribution~\cite{CMS:2021wlt}.
Bounds provided in the $s_H,m_{H_5}$ plane for the Georgi-Machacek model can approximately be translated to our parameter space by taking $s_H \propto \lambda v^2/M^2$.
They then cover the upper-left corner of \autoref{fig:quadruplets} and only become competitive with Higgs coupling measurements towards $\lambda\lesssim3$.
This would remain true until the end of the HL-LHC programme.
ATLAS obtains similar constraints in the fully leptonic $WZ$ channel by training an artificial neural network on eight variables including $m_{jj},\Delta\phi_{jj},\eta_V,H_T,E_T^\text{miss}$ to define signal and background regions before performing a likelihood fit on $m_{WZ}$~\cite{ATLAS:2022hnp, ATLAS:2022jho}.
A simple estimate for the sensitivity at FCC-hh can be obtained by using, as background, the SM vector-boson-fusion production of $WZ$ pairs for $m_{WZ}$ above the probed charged scalar mass, assuming an overall efficiency factor of $1\%$ (including leptonic branching ratios) for both signal and background.
Such a procedure reproduces the actual LHC sensitivity and, for the FCC-hh, indicates that the reach of single production is similar to that of pair production, in the region not already probed by Higgs coupling measurements at FCC-ee.

At future lepton colliders collecting 5\,ab$^{-1}$ at 3\,TeV or 10\,ab$^{-1}$ at 10\,TeV, the reach of single production estimated in the same way is comparable to that of the HL-LHC in the range of parameters shown in \autoref{fig:quadruplets}.
On the other hand, pair production can be expected to probe masses close to half of the centre-of-mass energy.

\subsubsection*{\texorpdfstring{$\delta_{h^3} /\delta_{VV}$}{delta h3 / delta VV } ratio}
\addcontentsline{toc}{subsubsection}{\texorpdfstring{$\quad\;\;\delta_{h^3} /\delta_{VV}$}{delta h3 / delta VV } ratio}

In summary, we have considered a model that is custodially symmetric up to small corrections, renormalisable, and generates only $\mathcal{O}_6$ at tree level and dimension-$6$.
It thus evades constraints on the $\widehat{T}$ parameter.
Single-Higgs couplings are moreover only modified at one-loop and dimension-$6$, or at tree-level and dimension-$8$.
The model also provides an opportunity to examine the $\delta_{h^3}/\delta_{VV}$ ratio quantitatively, since both corrections are calculable.
Putting aside vacuum stability considerations and including the sole $\hat{\Phi} H^3$ quartic potential term, we find
\begin{equation}
-\frac{\delta_{VV}}{\delta_{h^3}}
\;\: = \;\:
3\bigg(\frac{m_h}{4\pi v}\bigg)^2 +\bigg(\frac{m_h}{M}\bigg)^2
\;\:\approx\;\:
\frac{1}{200}+ \frac{1}{580}
\left(\frac{3 \text{ TeV}}{M} \right)^2
~,
\end{equation}
which is remarkably similar to the estimate of \autoref{eq:scaling}.
The structure of this explicit model is indeed such that it respects the power counting introduced in \autoref{sec:intro}.
The coupling of the quadruplet to the Higgs is a scalar quartic interaction, thus of coupling dimension 2, which is invariant under the $\lambda \to - \lambda$, $(\Phi,\widetilde{\Phi}) \to -(\Phi,\widetilde{\Phi})$ parity transformation.
Thus, after integrating out the heavy quadruplet, the coupling can only enter in the low energy EFT as $\kappa = \lambda^2$, which has coupling dimension 4.
On the one hand, this model demonstrates that UV completions with this power counting can exist and that we should thus be cautious about applying theory priors related to specific models in weighing the importance of future Higgs physics measurements.
On the other hand, the custodial quadruplet model seems to be the only working example at tree-level and is thus relatively unique.

The introduction of an additional $|\hat{\Phi}|^2 |H|^2$ quartic potential term to ensure vacuum stability extends the power counting discussed in \autoref{sec:intro}.
In particular, it allows for the presence of a $|H|^8$ operator at tree level in the EFT.
The perturbativity of such $|H|^8$ contribution needed for vacuum stability at small field values is less constraining than that of multi-boson scattering.
For perturbativity up to scales of the order of $6\,$TeV, the latter allows self-coupling modifications of order $|\delta_{h^3}|\approx 5$.
Although our study of the $\delta_{h^3}/\delta_{VV}$ ratio does not include the potential terms required for vacuum stability, one would expect that $\delta_{VV}$ constraints would be dominant in determining the range of allowed self-couplings.
Values of about $\delta_{h^3}\approx-400\%$ should still be consistent with HL-LHC prospects on $\delta_{VV}$, but an EFT treatment including only operators of lowest dimension starts breaking down in this regime.
At foreseeable future colliders, direct searches would push the quadruplet mass into the TeV region and self-coupling measurements would still probe untouched parameter space. Similar considerations would apply if the additional quartics $\hat{\Phi}^3 H$ and $|\hat{\Phi}|^4$ were included, leading to the appearance in the EFT of $|H|^{10}$ and $|H|^{12}$ operators at tree level.

\section{\texorpdfstring{$|\delta_{h^3}| \gg |\delta_{VV}|$}{delta h3 >> delta VV} for a pNGB Higgs: Gegenbauer potentials}
\label{sec:pNGB}
To further illustrate the points of \autoref{sec:intro}, we now examine the Higgs potential of the recently proposed `Gegenbauer Higgs' class of pNGB Higgs models~\cite{Durieux:2021riy,Durieux:2022sgm}.
In particular, we focus on the Gegenbauer's Twin model of \cite{Durieux:2022sgm}.
In this model, the Higgs is a pNGB of a spontaneous $\text{SO}(8)\to\text{SO}(7)$ global symmetry breaking at the scale $f$.
The block-diagonal $\text{SO}(4)$ subgroups of $\text{SO}(8)$ are gauged such that six of the Goldstone bosons are eaten, leaving only one pNGB to be identified as the SM Higgs.

The top sector Yukawas explicitly break the global symmetries, leading to an estimable contribution to the Higgs potential of the form \cite{Craig:2015pha,Barbieri:2015lqa,Low:2015nqa}
\beq
V_t \approx \frac{3 y_t^4 f^4}{64 \pi^2}  \bigg[ \sin^4 \frac{h}{f}\, \log \frac{a}{\sin^2 \frac{h}{f} } + \cos^4 \frac{h}{f}\, \log \frac{a}{\cos^2 \frac{h}{f} } \bigg] ~~,
\eeq
in the unitary gauge.
Here, $a$ is a dimensionless $\mathcal{O}(1)$ constant calculable within specific UV scenarios, which materialises the logarithmic-only dependence of the potential on coloured particle masses that is typical of Twin Higgs models.
In addition to this top-sector contribution and to the subdominant gauge one, the Gegenbauer's Twin setup assumes an extra source of explicit symmetry breaking in the UV which corresponds to a non-zero value for a spurion in the $n$-index tensor irreducible representation of the global $\text{SO}(8)$ symmetry, breaking it to $\text{SO}(4) \times \text{SO}(4)$.
This additional contribution to the scalar potential is radiatively stable, in that UV corrections at any loop order preserve its form.
As shown in \cite{Durieux:2022sgm}, its expression is
\beq
V_G = \epsilon f^4 G_n^{3/2} \left(\cos 2 h/f \right) ~~,
\label{eq:geg}
\eeq
where $G_n^{3/2}$ is the Gegenbauer polynomial of index $3/2$ and order $n$.
The purpose of this work is not to expound upon the details of the model, but instead to use it as an example scenario to investigate the physics of self-coupling corrections and other potential features in pNGB Higgs models.

\subsection{Effective field theory expansion}
\begin{figure}[t]
\centering
\adjustbox{width=\textwidth}{%
\includegraphics[width=0.455\textwidth]{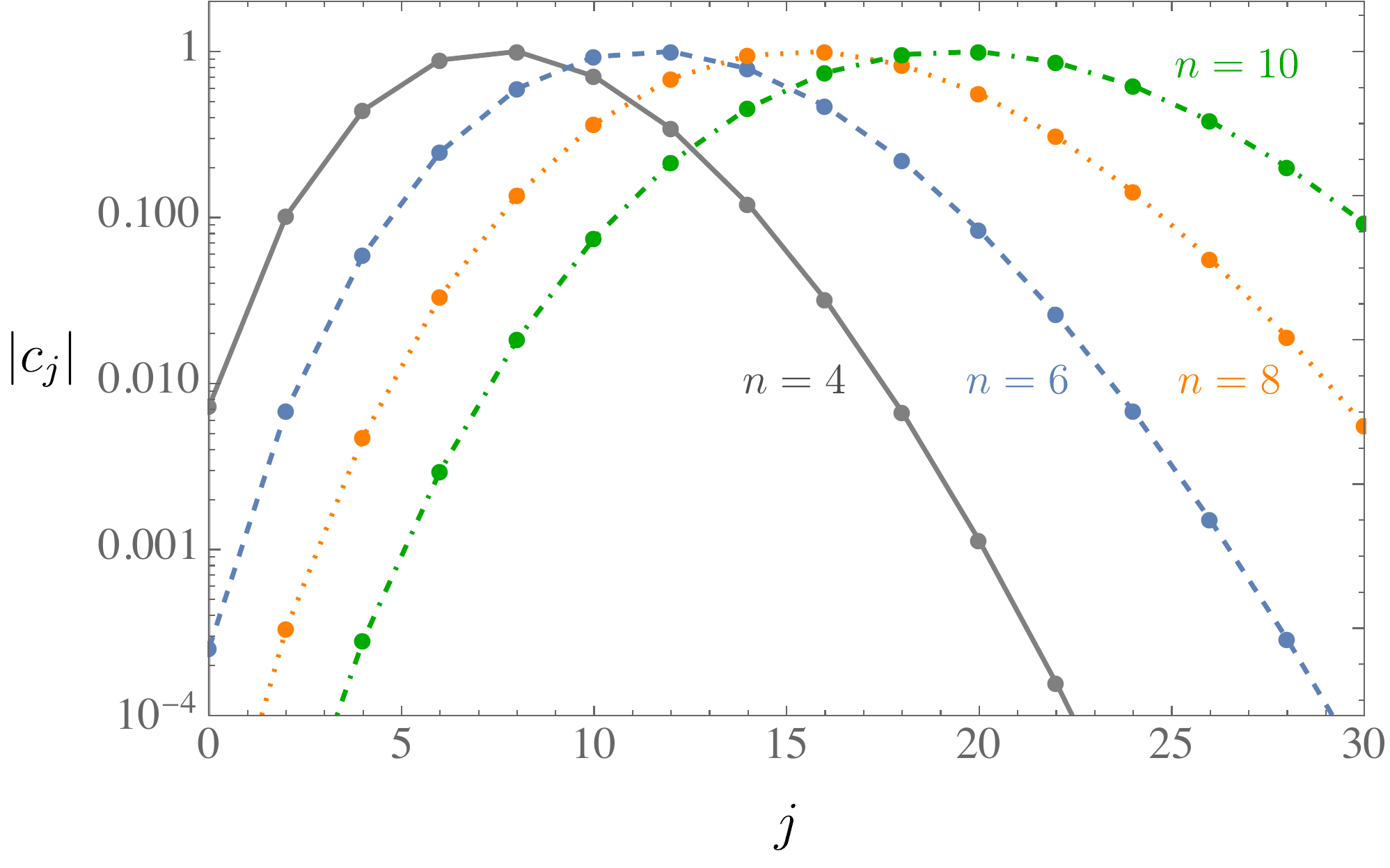} ~~ \includegraphics[width=0.45\textwidth]{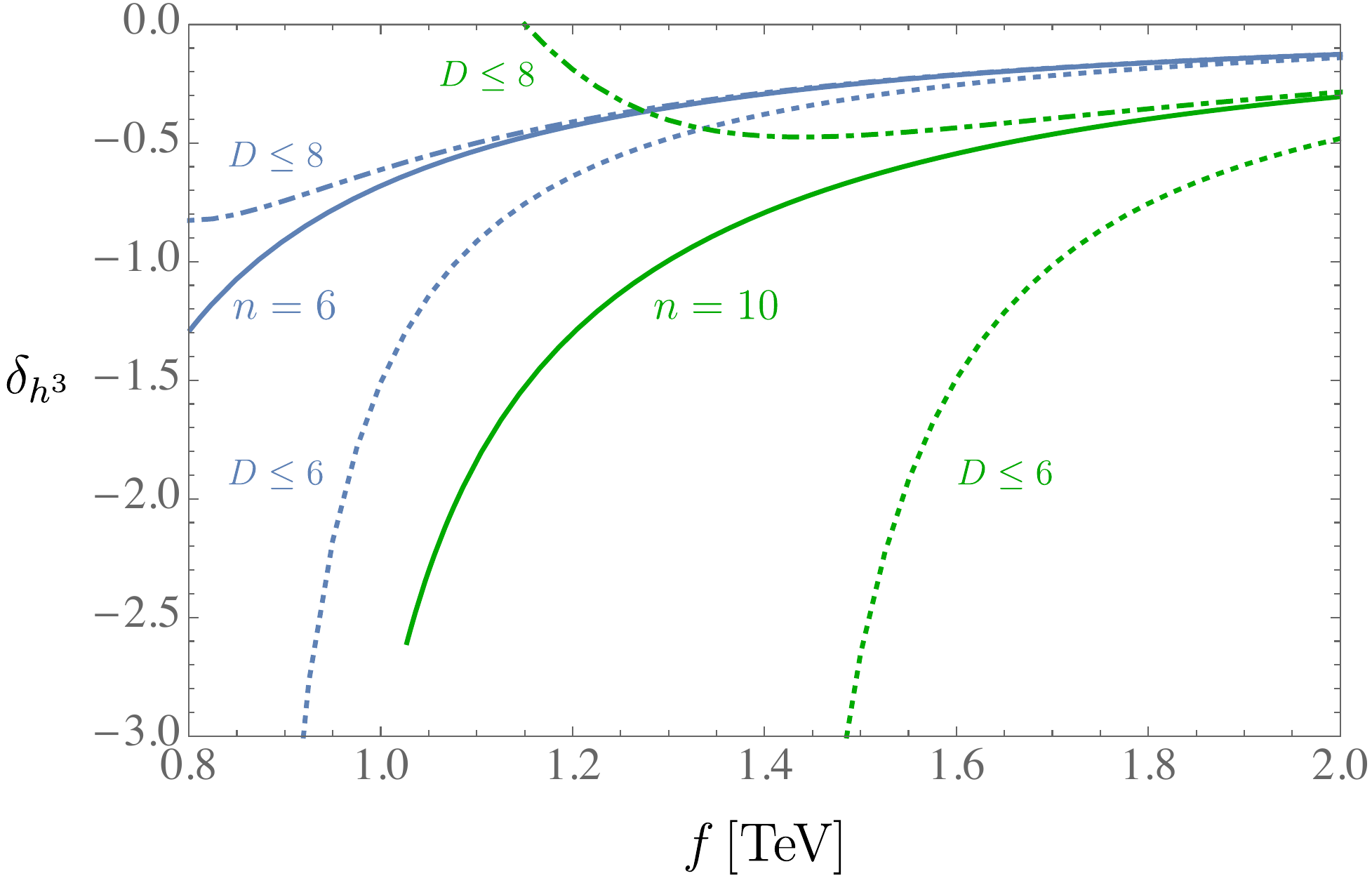}}
\caption{{\it Left:} The magnitude of the coefficients in the expansion of the Gegenbauer potential, \autoref{eq:expansion}, for various choices of $n$.
{\it Right:} Corrections to the Higgs self-coupling for $n=6$ and $10$ (solid lines), as compared to the results found from dimension-6 (dashed) and dimension-8 (dot-dashed) truncations of the full potential.}
\label{fig:traj}
\end{figure}

To commence, it is enlightening to investigate the accuracy of the EFT description of the Higgs potential, to understand in particular how well a truncation at dimension-6 performs.
To this end, first consider the Taylor expansion of the Gegenbauer potential,
\beq
G_n^{3/2} \left(\cos 2 h/f \right) = \mathcal{N}_n \sum^{\infty}_{j \,=\, 0} c_j \left(\frac{h}{f} \right)^j ~~,
\label{eq:expansion}
\eeq
where we have chosen $\mathcal{N}_n$, the overall normalisation factor, such that max~$\{ c_j \} = 1$.
In the left panel of \autoref{fig:traj}, we show the magnitudes of the $c_j$ for $n$ between $4$ and $10$.
They grow rapidly, up to $j = 2 n$, after which they begin asymptotically decreasing, providing a concrete realisation of a theoretical possibility discussed in \autoref{sec:intro}.

As a result of this behaviour, one expects that a description truncated at dimension-6 should fail to accurately capture the local form of the Higgs potential and, in particular, the corrections to the Higgs self-coupling.
To quantify this, we show in the right panel of \autoref{fig:traj} the Higgs self-coupling modification, relative to the SM value, for $n=6$ and $n=10$.
In each case, the full potential $V=V_t+V_G$ is used (solid lines) and compared against the result found from the truncation, at dimension-6 and -8, of its $\text{SU}(2)_L$ gauge-invariant form in terms of the doublet $H$.
To this end, we map $\sin^2 h/f$ in the unitary gauge to $2|H|^2/f^2$.
Clearly, a low-dimension EFT truncation of the potential fails to accurately capture the nature of the Higgs self-coupling corrections, especially as $n$ is increased.
This implies that any analysis of vacuum stability or unitarity violation based on a dimension-6, and perhaps even dimension-8, truncation may be inaccurate when applied to concrete models.\footnote{These conclusions are not restricted to the specific model considered here.
For instance, for a single $\text{U}(1)$ pNGB with a small source of explicit breaking $\text{U}(1)\to \mathcal{Z}_q$, the relevant Wilson coefficients scale as $|c_j|=q^j/j!$.
Thus the pattern of rapid growth up to a maximum, as seen in \autoref{fig:traj}, followed by factorial decrease is also found for this textbook $\text{U}(1)$ pNGB example.}

\subsection{Perturbative unitarity}
Given that large Higgs self-coupling corrections are possible in this model, it is natural to consider the scale at which perturbative unitarity breaks down.
Furthermore, given that the operators with the largest coefficients have high dimensions (see left panel of \autoref{fig:traj}), one would expect that the most relevant constraints arise from high-multiplicity scattering.
In \cite{Chang:2019vez}, such higher-point scattering processes were considered in general terms, leading to their eq.~(2.13) which gives the highest possible energy consistent with perturbative unitarity.

\begin{figure}[t]
\centering
\includegraphics[width=0.58\textwidth]{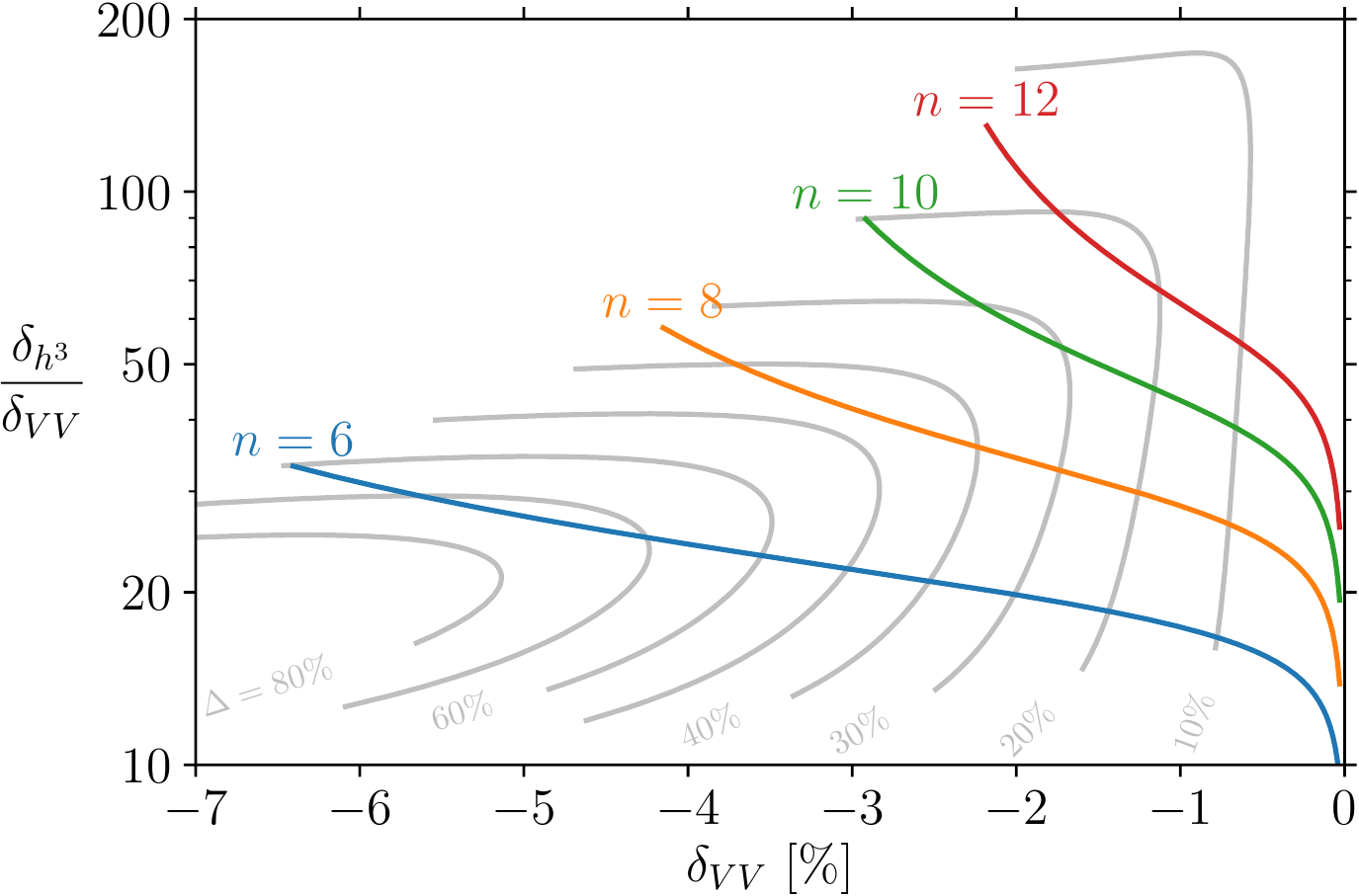}
\caption{The ratio of self-coupling to $hVV$ corrections in the Gegenbauer's Twin model of \cite{Durieux:2022sgm}, for several values of $n$.
In the empty upper-left corner of the figure, the SM vacuum is no longer the global minimum of the potential and therefore becomes unstable.
Grey contours show the fine-tuning of the Higgs mass and vacuum expectation value in this model.}
\label{fig:h3hvv}
\end{figure}

Applying that result to the Gegenbauer's Twin model, and focusing on processes involving Higgs bosons, we find that the lowest scale for perturbative unitarity breakdown is $E \gtrsim 6$ TeV for all of the parameter ranges considered and $n\geq 6$.
Furthermore, we find that for $6 \leq n \leq 12$ the strongest limit arises from $N$-point scattering with $10 \leq N \lesssim 40$.
Slightly lower scales would be obtained by including processes involving longitudinal $W,Z$ bosons~\cite{Chang:2019vez}.
While these limits are only approximate, as one is considering energies where higher-order perturbative contributions are becoming important, they nonetheless demonstrate that one may have a large energy range where the EFT description remains valid.
For comparison, the cross section for Higgs pair production used to extract $\delta_{h^3}$ is dominated by $m_{hh} \lesssim 1\;\mathrm{TeV}$ even at FCC-hh~\cite{Contino:2016spe}.
Depending on their exact nature, new states arising at about $6$~TeV could, however, be directly discovered at FCC-hh.

\subsection{\texorpdfstring{$\delta_{h^3} /\delta_{VV}$}{delta h3 / delta VV } ratio}
Finally, we consider the $\delta_{h^3}/\delta_{VV}$ ratio which is the focus of this paper.
In \autoref{fig:h3hvv}, we show this quantity for a range of single-Higgs coupling deviations $\delta_{VV}$, given by the standard pNGB Higgs expression $\sqrt{1-v^2/f^2}\,-1\,$, and a variety of $n$ values.
Contours of the fine-tuning of the Higgs mass and vacuum expectation value are also shown.
In parameter regions with negligible fine-tuning, self-coupling corrections can be an order of magnitude larger than single-Higgs coupling corrections, qualitatively consistent with the estimate in \autoref{eq:scaling}.
For instance, $\delta_{h^3}/\delta_{VV} \approx 30$ is obtained for $n=6$ and $\Delta\approx 60\%$.
In absolute terms, this corresponds to $\delta_{h^3}\approx -200\%$ and $\delta_{VV}\approx -6\%$ which is approximately matching current LHC constraints on single-Higgs couplings.
The same self-coupling deviation can be achieved together with $\delta_{VV}\approx -2.6\%$, which corresponds to the HL-LHC reach, for $n=10$ and a mild fine-tuning of $\Delta \approx 25\%$.
An even larger $\delta_{h^3}/\delta_{VV} \approx 75$ ratio is then obtained.
Pushing $\delta_{h^3}$ further requires higher $n$, increasing the fine-tuning, and is ultimately constrained by vacuum stability.
In summary, Gegenbauer's Twin provides yet another class of models for which Higgs self-coupling measurements probe otherwise untouched parameter space.

\section{Comparison to previous work}
\label{sec:others}
There is already a significant literature considering BSM modifications of the Higgs self-coupling, including \cite{Hollik:2001px,Kanemura:2002vm,Barger:2003rs,Grojean:2004xa,Giudice:2007fh,Gupta:2013zza,Efrati:2014uta,deBlas:2014mba,Azatov:2015oxa,Goertz:2015dba,Dawson:2015oha,Buttazzo:2015bka,Ginzburg:2015yva,Liu:2016idz,DiLuzio:2016sur,Baglio:2016bop,DiVita:2017eyz,DiLuzio:2017tfn,Carvalho:2017vnu,Chang:2019vez,Falkowski:2019tft,Agrawal:2019bpm,Abu-Ajamieh:2020yqi}.
Here, we briefly discuss those references which have compared single-Higgs to self-coupling deviations in specific scenarios.

Reference~\cite{Gupta:2013zza} considered Higgs self-coupling modifications in models with mixed-in singlet scalars, composite Higgs, and supersymmetric theories, concluding that in any scenario where nothing else is discovered at the HL-LHC the maximum deviations are likely to be at the $\sim 20\%$ level, much less than found for the models considered in this work.
This can be understood as follows.
The perturbative scenarios considered in \cite{Gupta:2013zza} do not have the same power-counting as the custodial quadruplet model, and hence display much more modest corrections.
The composite Higgs scenario assumes $\mathcal{O}(1)$ Wilson coefficients, in contrast with the Gegenbauer's Twin setup, again limiting the coupling corrections to moderate values.

Reference~\cite{DiLuzio:2017tfn} considered vacuum stability constraints, concluding that knowledge of the full set of higher-dimension operators is required in order to make a definitive statement.
Both models discussed here illustrate this important caveat well.
Perturbative unitarity constraints were considered, concluding that large self-coupling deviations are possible.
Explicit models were also examined; in particular, large self-coupling modifications were found when adding a scalar singlet to the SM.%
\footnote{The authors of \cite{DiVita:2017eyz} note that this comes at the additional cost of some fine-tuning in the Higgs quartic.
We have not commented on Higgs naturalness for the custodial quadruplet, since the Higgs bilinear and quartic are not strictly calculable quantities in this model.
In contrast to a singlet scalar, the quadruplet only induces, in general, contributions to the Higgs bilinear and quartic couplings at one $\hbar$ order higher than corrections to the Higgs self-coupling
(the $\hat{\Phi} H^3$ quartic term alone would only contribute to the Higgs bilinear at the two-loop level, as can also be inferred from the power counting of \autoref{sec:intro}).
The correlation between Higgs naturalness and self-coupling corrections is thus weaker than in the singlet model.}
Weak triplet and quadruplet models were also studied, with the conclusion that in these cases the self-coupling correction could be at most a few percent.
However, introducing only one electroweak multiplet at a time did not realise the custodial limit, in which significant self-coupling corrections can be achieved while remaining consistent with precision electroweak measurements.

Finally, \cite{Falkowski:2019tft} considered fine-tuning and vacuum stability aspects.
Vacuum stability constraints of the type of \autoref{eq:dh3-intervals-simple} were used to derive $|\delta_{h^3}|\lesssim 2$ for couplings of order $g_*=c_6^{1/4}=c_8^{1/6}\approx\pi$ and $\xi= (g_* v/M)^2\approx 0.1$ (so $M\approx 2.4\,$TeV).
In the custodial quadruplet model of \autoref{sec:tree}, larger couplings and smaller masses are however presently permitted since modifications of the single-Higgs couplings and $\widehat{T}$ parameter are further suppressed in the EFT and loop expansions, compared to the correction to the Higgs self-coupling.
In the Gegenbauer's Twin example of \autoref{sec:pNGB}, the Higgs potential receives sizeable contributions from operators of dimension higher than 8, which were neglected in the above vacuum stability considerations.
Our all-order analysis, however, shows that these do not allow to relax significantly fine-tuning and vacuum stability constraints on self-coupling modifications.

\section{Conclusions}
\label{sec:conclusions}
If the most readily accessible Higgs boson couplings are SM-like, then how non-standard could others, such as the self-coupling, reasonably be?
In this work, we have attempted to answer this question quantitatively.
We have found on general theoretical grounds that the ratio of BSM self-coupling modifications to vector coupling modifications, for generic UV completions that are not fine-tuned, will satisfy
\beq \label{eq:ratio_concl}
\left| \frac{\delta_{h^3}}{\delta_{VV}}  \right| \lesssim \min \left[ \left(\frac{4 \pi v}{m_h} \right)^2, \left(\frac{M}{m_h} \right)^2 \right] ~~,
\eeq
where $M$ is the lowest mass scale of new physics.
This formula is essentially a consequence of $\hbar$ counting in classes of models primarily generating the $|H|^6$ operator.
As a supporting proof-of-principle, we have presented a simple renormalisable extension of the SM by a custodial weak quadruplet, which saturates the ratio in \autoref{eq:ratio_concl} as a result of exhibiting the required power-counting in the microscopic interactions.
Furthermore, we have studied a pNGB-like Higgs scenario, Gegenbauer's Twin, which also comes close to saturating this ratio in addition to solving the naturalness problem of the SM.
Both of these models populate an important corner of theory space in which Higgs self-coupling modifications are dominant.

The punchlines for the experimental particle physics programme are simple.
Even though current LHC self-coupling analyses may appear weak as compared to more readily testable coupling deviations, for certain classes of models they are already probing parameter space that would otherwise be inaccessible at present.
For future programmes, this will remain true.
Even if a Higgs factory like the FCC-ee yields a $hZZ$ coupling measurement at two-sigma precision of $|\delta_{ZZ}| \lesssim 0.34\%$, the self-coupling deviations could in principle remain as large as $|\delta_{h^3}| \lesssim 200 \%$ according to \autoref{eq:scaling}, though in the models considered here we found slightly more moderate values, $|\delta_{h^3}| \lesssim 50 \%$.
Any measurement bettering this precision, whether at the $100\%$ level at the LHC or down to the $10\%$ level at future colliders, would be a highly valuable probe of unexplored microscopic Higgs terrain.
New physics could first manifest itself in Higgs self-coupling measurements.

We conclude by emphasising that, if a deviation in the self-coupling were indeed observed first, measuring the single-couplings would be of critical importance to characterise the new physics.
A global view of Higgs observables will be necessary in order to pin down the underlying BSM theory.
Given a deviation in $\delta_{h^3}$, our results may then be interpreted as providing a lower bound on $\delta_{VV}$ expected in generic models.

\acknowledgments

The authors would like to thank Aleksandr Azatov, Joan Elias Mir\'o, Adam Falkowski, Gian Giudice, Markus Luty, Michelangelo Mangano, David Marzocca, Filippo Sala, and Michael Trott for valuable discussions.
ES acknowledges partial support from the EU's Horizon 2020 programme under the MSCA grant agreement 860881-HIDDeN.

\appendix

\section{Custodial quadruplet decomposition} \label{sec:custodial}
To understand the decomposition of the custodially symmetric representations into electroweak quantum numbers, we work with the $\hat{\Phi} \sim (\mathbf{4},\mathbf{4})$ representation of $\text{SU}(2)_L \times \text{SU}(2)_R$.
The bifundamental Higgs field is written in terms of the doublet as
\begin{equation*}
\mathcal{H} = \begin{pmatrix} \epsilon H^\ast & H \end{pmatrix}
 =
\begin{pmatrix}
H^\ast_2 & H_1  \\
- H_1^\ast & H_2 
\end{pmatrix} ~~.
\end{equation*}
The Lagrangian for the custodial quadruplet is
\beq
\mathcal{L}_{\hat{\Phi}} = \mathcal{L}_{\rm kin}  - \frac{\lambda}{\sqrt{3}}\, \hat{\Phi}_{ijk}^{\;\;\;\;IJK} \mathcal{H}^{\ast i}_{\;\;\;I} \mathcal{H}^{\ast j}_{\;\;\;J} \mathcal{H}^{\ast k}_{\;\;\;K}~,
\eeq
where $\mathcal{L}_{\rm kin} = \tfrac{1}{2} \mathrm{Tr} [ D^\mu \hat{\Phi}^\ast D_\mu \hat{\Phi} - M^2 \hat{\Phi}^\ast \hat{\Phi} ] $ and $\mathrm{Tr}[\hat{\Phi}^\ast \hat{\Phi}] = \hat{\Phi}^{\ast\,ijk}_{\;\;\;\;\;IJK} \hat{\Phi}_{ijk}^{\;\;\;\;IJK} $.
The $\hat{\Phi}$ decomposes into two complex $\text{SU}(2)_L $ quadruplets, $\Phi \sim \mathbf{4}_{1/2}$ and $\widetilde{\Phi} \sim \mathbf{4}_{3/2}\,$, as
\begin{equation}
\begin{aligned}
\hat{\Phi}_{ijk}^{\;\;\;\;222} & =  \widetilde{\Phi}_{ijk}~,\\
 \hat{\Phi}_{ijk}^{\;\;\;\;221} & =   \hat{\Phi}_{ijk}^{\;\;\;\;212} =  \hat{\Phi}_{ijk}^{\;\;\;\;122} =  \Phi_{ijk} / \sqrt{3}~,\\
\hat{\Phi}_{ijk}^{\;\;\;\;112} & =   \hat{\Phi}_{ijk}^{\;\;\;\;121} =  \hat{\Phi}_{ijk}^{\;\;\;\;211} = -\,  \epsilon_{ia} \epsilon_{jb} \epsilon_{kc} \Phi^{\ast \,abc} / \sqrt{3}~,\\
 \hat{\Phi}_{ijk}^{\;\;\;\;111} & =  \epsilon_{ia} \epsilon_{jb} \epsilon_{kc} \widetilde{\Phi}^{\ast\, abc}~,
\end{aligned}
\end{equation}
where $\Phi, \widetilde{\Phi}$ are canonically normalised.
Using $\mathcal{H}^{i}_{\;\;1} = (\epsilon H^\ast)^i$ and $\mathcal{H}^{i}_{\;\;2} = H^{i}$ we see that the custodially symmetric interaction decomposes as
\bea
 - \lambda \Big( H^\ast H^\ast (\epsilon H) \Phi  + \frac{1}{\sqrt{3}} H^\ast H^\ast H^\ast  \widetilde{\Phi}\Big) + \mathrm{h.c.},
\eea
where $\text{SU}(2)_L$ indices are suppressed.
Thus, the full $\text{SU}(2)_L \times \text{SU}(2)_R$ symmetry enforces the couplings in \autoref{eq:custodialcoup}.

\section{Vacuum stability at dimension-8}
\label{sec:vacsta}

Following~\cite{Falkowski:2019tft}, we consider vacuum stability at small field values in an EFT featuring only the dimension-6 and -8 contributions to the Higgs potential, \autoref{eq:c6c8}.
Including operators of even higher dimension would not qualitatively affect the implications of vacuum stability on perturbative unitarity and the convergence of the EFT expansion.

Similarly to \autoref{eq:delta-def}, let us define
\begin{equation}
\delta_{h^4} \equiv \frac{C_{h^4}-C^{\text{SM}}_{h^4}}{C^{\text{SM}}_{h^4}}
\end{equation}
from the coefficient of the on-shell, momentum-independent, four-Higgs amplitude.
Qualitatively, vacuum stability at small field values demands that a large $\delta_{h^3}$ be compensated by an even larger $\delta_{h^4}$.
Including only $|H|^6$ and $|H|^8$ operators, the exact vacuum stability requirements are:
\begin{equation}
\delta_{h^4} \ge
	\begin{cases}
	6\,  \delta_{h^3}	& \text{for }\delta_{h^3}\in[0,2]~~,\\
	14\, \delta_{h^3} - 16	& \text{for }\delta_{h^3}\in[2,4]~~,\\
	6\,  \delta_{h^3} + (\delta_{h^3})^2	& \text{elsewhere}~~.
	\end{cases}
\label{eq:stability-cond}
\end{equation}%
In turn, this demands a minimal contribution to $\delta_{h^3}$ from the dimension-8 operator $|H|^8$.
Recalling the decomposition of \autoref{eq:dh3-decomposition} and defining the ratio $\rat\equiv \delta_{h^3}^{(8)} / \delta_{h^3}^{(6)}$, one obtains that
\begin{equation}
\delta_{h^3}^{(6)} \le
	\begin{cases}
	8/(3\rat + 4)		&\text{for }\rat\ge 0~~,\\
	2\rat/(1+\rat)^2	&\text{for }-1<\rat<0~~.
	\end{cases}
\end{equation}
The second inequality is equivalent to \autoref{eq:limit-on-negative-r}.
For $\delta_{h^3}^{(6)}\in [-4,16/11]$, vacuum stability at small field values does therefore not require the relative contribution of the dimension-8 operator $|H|^8$ to be larger than $50\%$, i.e.\ $|\rat|\le 1/2$.

The vacuum stability requirements of \autoref{eq:stability-cond} imply allowed intervals on $\delta_{h^3}$
\begin{equation}
\begin{aligned}
		\begin{cases}
		1-\sqrt{ 1- 2\delta_{h^3}^{(6)} }	&\text{for } \delta_{h^3}^{(6)}<0	\\
		\delta_{h^3}^{(6)}			&\text{for } \delta_{h^3}^{(6)}\in[0,2]
		\end{cases}
	&\;\le\delta_{h^3}\le\;
		\begin{cases}
		1+\sqrt{ 1- 2\delta_{h^3}^{(6)} }	&\text{for }\delta_{h^3}^{(6)}<-4\\
		(8-\delta_{h^3}^{(6)})/3		&\text{for }\delta_{h^3}^{(6)}\in[-4,2]
		\end{cases}
,\\
		-\sqrt{2\delta_{h^3}^{(8)}}
	&\;\le\delta_{h^3}\le\;
		\begin{cases}
		\delta_{h^3}^{(8)}/4+2		&\text{for }\delta_{h^3}^{(8)}\in[0,4]\\
		\sqrt{2\delta_{h^3}^{(8)}}	&\text{for }\delta_{h^3}^{(8)}>4
		\end{cases}
,
\end{aligned}
\label{eq:dh3-intervals}
\end{equation}
which depend on $c_6$, $c_8$, and $M$ through the combinations
\begin{equation}
\begin{aligned}
\delta_{h^3}^{(6)} &= \frac{2 c_6 v^4}{M^2m_h^2}
	\approx \frac{c_6}{(4\pi)^4} \left(\frac{110\,\text{TeV}}{M}\right)^2
	~~,\\
\delta_{h^3}^{(8)} &= \frac{4 c_8 v^6}{M^4m_h^2}
	\approx \frac{c_8}{(4\pi)^6} \left(\frac{22\,\text{TeV}}{M}\right)^4
	~~.
\end{aligned}
\end{equation}
In particular, this leads to \autoref{eq:dh3-intervals-simple}.

\end{fmffile}
\bibliographystyle{apsrev4-1_title}
\bibliography{biblio}

\end{document}